\documentclass[11pt]{amsart}  
\usepackage[left=1.3in,right=1.3in,bottom=1.3in]{geometry}  
\usepackage[T1]{fontenc}

\usepackage{amssymb,amsmath,amscd}
\usepackage{amsthm}
\usepackage[parfill]{parskip}
\usepackage{upquote}

\usepackage{upgreek}
\usepackage{mathrsfs}
\usepackage{latexsym}
\usepackage{enumerate}
\usepackage{graphicx} 
\usepackage{tikz-cd}
\usepackage{galois}
\usepackage{stmaryrd}
\usepackage[boxed,linesnumbered,ruled,vlined]{algorithm2e}
\usepackage{todonotes}

\theoremstyle{plain}
\newtheorem{theorem}[]{Theorem}
\newtheorem{lemma}[theorem]{Lemma}
\newtheorem{proposition}[theorem]{Proposition}
\newtheorem{corollary}[theorem]{Corollary}

\theoremstyle{definition}
\newtheorem{definition}[]{Definition}
\newtheorem{example}[]{Example}
\newtheorem{remark}[]{Remark}

\newcommand{\id}{\textsc{id}} 
\newcommand{\Fix}{\textsc{Fix}} 
\newcommand{\LLN}{\mathcal{L}} 
\newcommand{\conv}{\chi} 
\newcommand{\inagg}{\alpha} 
\newcommand{\dist}{\delta} 
\newcommand{\outagg}{\omega} 
\newcommand{\liab}{\ell} 
\newcommand{\epymt}{p} 
\newcommand{\trans}{\Xi} 

\newcommand{\pow}[1]{\mathbb{P}\left(#1\right)}  
\newcommand{\permset}{W}
\newcommand{\perm}{w}

\begin{document}

\title{Clearing Sections of Lattice Liability Networks}
\author{Robert Ghrist}
\author{Julian Gould}
\author{Miguel Lopez}
\author{Hans Riess}
\address{Department of Mathematics and Electrical \& Systems Engineering \\
         University of Pennsylvania \\
         Philadelphia, PA}
\address{Department of Mathematics \\
         University of Pennsylvania \\
         Philadelphia, PA}
\address{Applied Mathematics \& Computational Science \\
         University of Pennsylvania \\
         Philadelphia, PA}
\address{School of Electrical and Computer Engineering \\
         Georgia Institute of Technology \\
         Atlanta, GA}

\begin{abstract}
Modern financial networks involve complex obligations that transcend simple monetary debts: multiple currencies, prioritized claims, supply chain dependencies, and more. We present a mathematical framework that unifies and extends these scenarios by recasting the classical Eisenberg-Noe model of financial clearing in terms of lattice liability networks. Each node in the network carries a complete lattice of possible states, while edges encode nominal liabilities. Our framework generalizes the scalar-valued clearing vectors of the classical model to lattice-valued clearing sections, preserving the elegant fixed-point structure while dramatically expanding its descriptive power. Our main theorem establishes that such networks possess clearing sections that themselves form a complete lattice under the product order. This structure theorem enables tractable analysis of equilibria in diverse domains, including multi-currency financial systems, decentralized finance with automated market makers, supply chains with resource transformation, and permission networks with complex authorization structures. We further extend our framework to chain-complete lattices for term structure models and multivalued mappings for complex negotiation systems. Our results demonstrate how lattice theory provides a natural language for understanding complex network dynamics across multiple domains, creating a unified mathematical foundation for analyzing systemic risk, resource allocation, and network stability.
\end{abstract}


\maketitle

\section{Introduction}
\label{sec:intro}

The study of financial networks has emerged as a crucial framework for understanding systemic risk and economic stability. Since the 2008 financial crisis, there has been growing recognition that the interconnected nature of financial institutions can amplify local disturbances into system-wide catastrophes. A landmark contribution to this analysis was the Eisenberg-Noe model \cite{eisenberg2001systemic}, which provided a rigorous mathematical framework for analyzing clearing payments in a network of financial obligations. This model demonstrated how lattice-theoretic methods, particularly the Tarski Fixed Point Theorem, could guarantee the existence of clearing payment vectors in financial networks.

While groundbreaking, the classical Eisenberg-Noe framework is limited to scalar-valued payments between institutions. Modern financial networks often involve more complex state spaces: payments in multiple currencies, obligations with different seniorities, or resources that cannot be directly compared. Supply chains, for instance, may need to track both material flows and priority signals. Social networks might need to model both information spread and trust metrics. This suggests the need for a more general framework that can handle payments and obligations taking values in arbitrary lattices while preserving the elegant fixed-point structure of the original model.

Our work provides such a generalization by recasting the clearing problem in the language of order lattices, monotone maps, and fixed point theory: see the Appendix for background. The key innovation is our concept of a \emph{lattice liability network} -- a type of data structure over a quiver where each vertex carries a complete lattice of possible payments and each edge specifies a nominal liability. Nodes also carry pay-in and pay-out aggregators and distributors that manage resource flows while respecting lattice structure. This framework naturally accommodates:
\begin{enumerate}
\item Multiple types of obligations (currencies, commodities, etc.) with conversion and transformation;
\item Priority structures and partial orders on payments;
\item Nonlinear aggregation rules for incoming and outgoing resources;
\item Scheduled payment plans with present-value discounting.
\end{enumerate}

The main result -- Theorem \ref{thm:existence} -- establishes that under appropriate monotonicity conditions, such networks always possess clearing sections as assignments of payment lattice elements to vertices that simultaneously satisfy all local constraints and global consistency conditions. Moreover, the set of all clearing sections forms a complete lattice, generalizing the order structure found in the classical Eisenberg-Noe model. This result provides a unified framework for analyzing equilibria in a broad class of network models while preserving the computational advantages of lattice-theoretic methods.

We demonstrate the flexibility of our framework through several applications. These include multi-currency financial networks, decentralized finance with automated market makers, supply chains with resource transformation, and permission networks with complex authorization structures. We extend our framework to chain-complete lattices to handle term structure models and develop multivalued extensions for complex negotiation systems. In each case, the lattice structure provides both theoretical guarantees about the existence of clearing solutions and practical insights about system behavior.

The paper is organized as follows. Section \ref{sec:EN} reviews the classical Eisenberg-Noe model and its lattice-theoretic foundations. Section \ref{sec:motivations} discusses the mathematical motivation for generalizing beyond scalar-valued models. Section \ref{sec:LLN} introduces our general framework of lattice liability networks. Sections \ref{sec:clearing}-\ref{sec:MAIN-THM} define clearing sections and exogenous resources, and present our main existence theorem for clearing sections. Section \ref{sec:computational} develops a distributed algorithm for computing clearing sections. The remaining sections explore various applications and extensions, including the classical Eisenberg-Noe model as a special case (\S\ref{sec:EN-LLN}), multi-currency systems (\S\ref{sec:currencies}), decentralized finance (\S\ref{sec:amm}), supply chains (\S\ref{sec:supply}), residuated lattices for fuzzy payments (\S\ref{sec:fuzzy}), concrete multi-attribute network applications (\S\ref{sec:applications}), permission networks (\S\ref{sec:permission}), extension to chain-complete lattices (\S\ref{sec:chaincomplete}), term structure models (\S\ref{sec:termstructure}), multivalued extensions (\S\ref{sec:multival}), and negotiation networks (\S\ref{sec:negotiation}). Appendices contain basic background material on lattices, fixed point theorems, chain-complete lattices, and residuated lattices.

Throughout, we emphasize how the interplay between order theory and network structure provides a powerful lens for understanding complex systems. Our framework shows that many seemingly different network phenomena -- from financial contagion to supply chain dynamics -- share a common mathematical core in the form of lattice-valued network flows and their fixed points.

\section{The Eisenberg-Noe Model}
\label{sec:EN}

The concept of clearing vectors, formalized by Eisenberg and Noe, has become fundamental in the analysis of interconnected financial systems. Consider a directed network $(V,E)$ with vertex set $V = \{1,\ldots,n\}$ representing financial institutions. An edge $e \in E$ from vertex $i$ to vertex $j$ represents a nominal liability (debt obligation) $\liab_e$ from institution $i$ to institution $j$.

For each vertex $v \in V$, let $\iota_v$ denote its external (or \emph{exogenous}) assets. The \emph{total liability} at $v$ equals $\bar{\liab}_v=\sum_{e \in s^{-1}(v)} \liab_e$. For each edge $e \in E$, define the proportional factor
\[
\pi_e 
= \frac{\liab_e}{\bar{\liab}_{s(e)}}
= \frac{\liab_e}{\displaystyle \sum_{e' : s(e')=s(e)} \liab_{e'}}
\]
representing the fraction of vertex $s(e)$'s total obligations owed along edge $e$.

A clearing vector $\mathbf{x} = (x_v)_{v \in V}$ represents the actual payments made by each institution, satisfying two key principles:
\begin{enumerate}
\item Limited liability: No institution pays more than its total obligations.
\item Absolute priority: Institutions pay out as much as possible to creditors before retaining any value.
\end{enumerate}

Formally, for each vertex $v \in V$, we require:
\begin{equation}
\label{eq:EN}
x_v 
= 
\min\left\{
\sum_{e \in s^{-1}(v)} \liab_e
\; ,\; 
\iota_v \; + \!\sum_{e \in t^{-1}(v)} \pi_e x_{s(e)}
\right\}
\end{equation}

\begin{figure}
    \centering
    \includegraphics[width=4.5in]{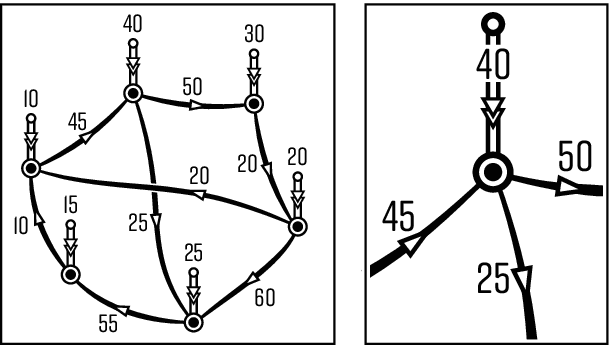}
    \caption{\small A classical Eisenberg-Noe system with six financial institutions, each with exogenous resources [left]. Liabilities are indicated along directed edges. At each node [right], a clearing section must balance incoming payments and resources with outgoing liabilities. At this node, pay-outs are scaled 2:1 to the two debtors, based on proportional liabilities. Both debts here can be paid in full assuming it receives what it is owed. However, elsewhere in the system (the right and leftmost nodes), not all debts can be paid, and defaults spread through the network, including to this node. Clearing vectors represent equilibrium payouts which balance pay-ins and pay-outs.}
\label{fig:EN}
\end{figure}

This can be formulated as a fixed point problem. 
Let $C^0 = \prod_{v \in V} [0,\bar{\liab}_v]$ denote the product of intervals, where $\bar{\liab}_v = \sum_{e \in s^{-1}(v)} \liab_e$ is the total liability of vertex $v$. This is a complete lattice under the product order. Define $\Phi: C^0 \to C^0$ by:
\begin{equation}
\label{eq:EN-Phi}
\Phi_v(\mathbf{x}) 
= 
\min\left\{
\sum_{e \in s^{-1}(v)} \liab_e
\; , \; 
\iota_v \; + \! \sum_{e \in t^{-1}(v)} \pi_e x_{s(e)}
\right\}
\end{equation}

A clearing vector is then a fixed point of $\Phi$. The mapping $\Phi$ is monotone on the complete lattice $C^0$, and thus by Tarski's Fixed Point Theorem, a greatest clearing vector exists. Under mild regularity conditions (such as the presence of positive external assets), this maximal clearing vector is unique.

The Eisenberg-Noe model has profoundly influenced financial network theory and practice. Its theoretical contribution lies in formalizing contagion processes through the elegant application of fixed-point methods to payment networks \cite{elsinger2006risk,cifuentes2005liquidity}. By proving the existence and uniqueness of clearing vectors under specific conditions, it established a rigorous mathematical foundation for systemic risk assessment \cite{acemoglu2015systemic}.

From a regulatory perspective, this framework has provided authorities with quantitative tools to evaluate financial stability, informing policy decisions on capital requirements \cite{gai2011complexity}, liquidity standards \cite{arnold2012systemic}, and the designation of systemically important financial institutions \cite{glasserman2015likely}. Central banks and regulatory bodies worldwide have incorporated network models inspired by Eisenberg-Noe into their stress-testing frameworks \cite{upper2011simulation,cont2013network}.

The model has generated numerous extensions across several dimensions:
\begin{enumerate}
\item \textit{Default mechanisms}: Rogers and Veraart \cite{rogers2013failure} incorporated bankruptcy costs that reduce recovery values during default, while Elsinger \cite{elsinger2009financial} introduced seniority structures in debt obligations.

\item \textit{Cross-holdings and overlapping portfolios}: Elliott et al.~\cite{elliott2014financial} extended the framework to include equity cross-holdings, while Caccioli et al.~\cite{caccioli2014stability} and Cont and Schaanning \cite{cont2019monitoring} modeled contagion through common asset holdings and fire sales.

\item \textit{Dynamic extensions}: Capponi and Chen \cite{capponi2015systemic} developed multi-period models accounting for strategic behavior and uncertainty in future asset values. Barratt and Boyd \cite{barratt2020multi} formulated liability clearing as a convex optimal control problem over multiple time periods with realistic financing constraints. Banerjee et al.~\cite{banerjee2018uniqueness} extended the framework to dynamic settings.

\item \textit{Multi-layered complexity}: Kusnetsov and Veraart \cite{kusnetsov2019interbank} incorporated multiple maturities and insolvency law, Feinstein \cite{feinstein2019obligations} extended the model to multiple assets and physical delivery, and Banerjee and Feinstein \cite{banerjee2019impact} incorporated contingent payments.

\item \textit{Continuous-time extensions}: Bardoscia et al. \cite{bardoscia2019full} proposed a full payment algorithm, while other researchers have developed continuous-time frameworks \cite{banerjee2019impact}.

\item \textit{Incomplete information}: Eisenberg and Noe's original assumption of complete knowledge has been relaxed by Anand et al.~\cite{anand2015filling} and Gandy and Veraart \cite{gandy2017bayesian}, who proposed methods for network reconstruction from partial information.
\end{enumerate}

Empirical applications of the model have illuminated real-world interbank dependencies and policy implications \cite{cont2013network,furfine2003interbank,mistrulli2011assessing}. These studies have demonstrated both the value of network-based approaches and the challenges in obtaining comprehensive data on financial interconnections.

Despite these advances, the classical Eisenberg-Noe framework has important limitations. It assumes a static network structure and deterministic asset values, when in reality financial networks evolve dynamically \cite{barucca2016network} and asset values fluctuate stochastically \cite{glasserman2016contagion}. Its restriction to scalar-valued payments cannot adequately represent complex financial instruments with multiple attributes, state-contingent payoffs \cite{demange2018multilayer}, or contracts denominated in different currencies. Furthermore, the original framework does not account for strategic behavior among financial institutions \cite{jackson2015networks}, heterogeneous preferences over payment timing \cite{biais2012dynamic}, or non-monetary obligations.

A significant innovation was proposed by Barratt and Boyd \cite{barratt2020multi}, who reconceptualized the problem as determining a sequence of payments over multiple time periods using convex optimal control. Their approach differs from the original Eisenberg-Noe framework in two crucial aspects. First, they model liability clearing as a multi-period process, whereas Eisenberg-Noe determines a single set of payments. Second, they introduce a realistic financing constraint that entities cannot pay more than the cash they have on hand, which implies that multiple steps may be needed to clear liabilities. Their formulation also allows for various objective functions and constraints, making it more flexible for real-world applications.

Our work addresses these limitations by recasting the clearing problem in the language of lattice liability networks. This approach enables us to model broader classes of financial relationships including: multi-layer obligations \cite{demange2018multilayer}, seniority structures \cite{elsinger2009financial,fischer2014no}, 
state-contingent contracts and complex derivatives \cite{duffie2015systemic,amini2016risk}, or 
supply chains  \cite{acemoglu2012network,baqaee2019macroeconomic}.

By generalizing from real-valued payments to payments taking values in arbitrary complete lattices, our framework preserves the mathematical elegance of the original Eisenberg-Noe model while substantially expanding its descriptive power and application domains.

\section{Mathematical Motivations}
\label{sec:motivations}

The Eisenberg-Noe model's use of order theory and the Tarski Fixed Point Theorem was remarkably prescient. While most early work on financial networks relied on linear programming and classical optimization techniques \cite{elsinger2006risk}, E-N's order-theoretic approach revealed deeper structural properties of clearing systems. Their key insight was that monotonicity --- the principle that more incoming payments enables more outgoing payments --- is fundamentally independent of linearity or continuity.

In the years following Eisenberg and Noe's seminal work, subsequent research largely returned to classical optimization methods. Various works reformulated clearing problems as linear programs \cite{cifuentes2005liquidity} or applied variational techniques \cite{acemoglu2015systemic}. These yield powerful computational tools but obscure the underlying structure that makes clearing solutions robust. Indeed, the advantages of lattice methods become apparent precisely when classical smoothness fails: bankruptcy costs introduce discontinuities \cite{rogers2013failure}, currency slippage breaks linearity, and strategic defaults create non-convexities \cite{banerjee2018uniqueness}.

This suggests reconsidering financial networks through a purely order-theoretic lens. The classical E-N model implicitly uses the simplest possible lattices --- closed intervals in $\mathbb{R}$ representing payment amounts. But many modern financial relationships involve more complex ordered structures:
priority classes of debt forming non-total orders \cite{elsinger2009financial}; 
contingent obligations under different scenarios \cite{demange2018multilayer}; and
fuzzy or probabilistic commitments to pay \cite{gandy2017bayesian,deMarco2020fictitious}. Even beyond finance, many network phenomena naturally involve lattice-valued states: production capabilities in supply chains \cite{acemoglu2012network}, trust metrics in social networks \cite{jackson2015networks}, or resource constraints in distribution systems \cite{baqaee2019macroeconomic}.

The challenge is finding the right mathematical framework to handle such generalized network states while preserving E-N's fundamental insight about monotone clearing. Several modern approaches suggest themselves. Network sheaf theory provides a powerful language for local-to-global consistency problems, especially via the sheaf [Hodge] Laplacian \cite{hansen2019,hansen2020,Ghrist_2022}.  However, the sheaf-theoretic perspective struggles with the directed cycles inherent in financial obligations. The existence of feedback loops --- where bank A owes bank B who owes bank C who owes bank A --- breaks the gradient-like dynamics of a sheaf Laplacian. The quiver Laplacian of Sumray, Harrington, and Nanda is more promising, but does not truly encode cyclic dynamics, being a one-sided zig-zag sheaf Laplacian \cite{sumray2024quiver}.

Our key insight is that while the network sheaf approach fails for cyclic financial networks, the order-theoretic core of Tarski's Fixed Point Theorem remains valid. This suggests recasting the entire framework in terms of a more general sheaf-like data structure. In particular, we adopt the term ``clearing section'' to emphasize the parallel with global sections of a sheaf: both represent globally consistent assignments that satisfy local constraints. Just as a global section of a sheaf assigns compatible data to each open set, a clearing section assigns compatible payment values to each vertex in our network.

This reformulation preserves the essential features of the E-N model while dramatically expanding its expressive power. The lattice structure at each vertex can encode complex local constraints, while monotone maps along edges ensure that the network's dynamics respect these ordered relationships. Most crucially, Tarski's theorem continues to guarantee the existence of clearing sections even in settings where classical fixed point theorems (requiring continuity or contraction) would fail.

The quiver representation framework offers several additional advantages. First, it cleanly separates the network topology (encoded in the quiver) from the algebraic structure at each node (the choice of lattice and monotone maps). This modularity simplifies both theoretical analysis and practical implementation. Second, it suggests natural generalizations to multi-valued relationships \cite{Zhou1994} and time-varying networks \cite{barucca2016network}. Finally, it connects financial network theory to a rich mathematical literature on representations of directed graphs, opening new avenues for structural analysis.

The rest of this paper develops this framework rigorously. We begin in \S\ref{sec:LLN} with the formal definition of lattice liability networks as quiver representations in the category of complete lattices and monotone maps. Technical background on lattices, Tarski's theorem, and quiver representations is provided in the appendices.

\section{Lattice Liability Networks}
\label{sec:LLN}

To reformulate the classical Eisenberg-Noe model into a novel data structure, we begin at the bottom, with the network. Classical and modern financial systems alike are built upon directed networks. For maximum precision, we borrow the notion of a \emph{quiver}, whose uses in everything from representation theory to data science are well-established \cite{keller2017quiver,bourget2023geometry,sumray2024quiver}.

\begin{definition}
\label{def:quiver}
A \emph{quiver} $Q = (Q_0,Q_1,s,t)$ consists of a set $Q_0$ of vertices, a set $Q_1$ of arrows, and maps $s,t: Q_1 \to Q_0$ specifying the source and target of each arrow. 
\end{definition}

Under this notation, $t^{-1}(v)$ consists of edges pointing into $v$ and $s^{-1}(v)$ consists of the edges pointing out of $v$. This combinatorial structure permits multiple edges between a pair of vertices, as well as self-loops. This will be useful in managing external resources (see \S\ref{sec:exo}).  No deep results from quiver theory are needed for the remainder of this work.

The E-N model derives much of its elegance from working with scalar-valued payments. However, many real-world financial and economic networks involve more complex state spaces: payments in multiple currencies, obligations with different seniorities, or resources that cannot be directly compared. This suggests generalizing from real-valued payments to payments taking values in complete lattices.

When liabilities and payments are real numbers, addition aggregates incoming flows, and multiplication by proportional factors distributes outgoing payments. When multiple incoming resources of different types arrive at a node, we need a systematic way to aggregate them. When a node must distribute its resources among various obligations, we need a mechanism to ensure conservation of resources while respecting priorities.

To handle these complexities systematically, we introduce the \emph{lattice liability network} as a data structure over a quiver. Each vertex carries a complete lattice of possible states, encoding the allowable configurations at that node. The edges carry nominal liabilities (to bound the required payments). Most subtly, each vertex requires both aggregation operators to collect incoming resources and a distribution mechanism to allocate outgoing payments. 

\begin{definition}[Lattice Liability Network]
\label{def:LLN}
Let $Q=(V,E,s,t)$ be a finite quiver. A \emph{lattice liability network} $\LLN_Q$ consists of:
\begin{enumerate}
  \item For each vertex $v\in V$, a complete \emph{payment lattice} $L_v$
  \item For each edge $e\in E$, a \emph{nominal liability} 
    \[
    \liab_e\in L_{s(e)}
    \]
  \item For each vertex $v\in V$, a \emph{pay-in aggregator}
    \[
    \inagg_v: \prod_{e\in t^{-1}(v)} L_{s(e)} \longrightarrow L_v
    \]
    which is monotone in each coordinate
  \item For each vertex $v\in V$, a \emph{distribution-aggregation pair} $(\dist_v, \outagg_v)$ consisting of a monotone \emph{distributor} $\dist_v$ and monotone \emph{pay-out aggregator} $\outagg_v$,
    \[
    \dist_v: L_v \longrightarrow L_v^{\,|s^{-1}(v)|}
    \quad : \quad
    \outagg_v: L_v^{\,|s^{-1}(v)|} \to L_v
    \quad : \quad 
    \outagg_v\circ\dist_v=\id_{L_v}
    \]
    with the liability bound for each $e\in s^{-1}(v)$:
    \begin{equation}
        \label{eq:Bounded-Liability}
    \left[\dist_v(\top_v)\right]_e \leq \liab_e ,
    \end{equation}
    where $\top\in L_v$ is the maximal element.
\end{enumerate}
\end{definition}

The pay-in aggregator $\inagg_v$ performs two conceptually distinct operations: it both converts incoming quantities from their source lattices $L_{s(e)}$ into the local lattice $L_v$, and combines these converted quantities. This eliminates the need for separate conversion operators while preserving the model's ability to handle heterogeneous resource types. 

The pay-out aggregator acts in conjunction with the distributor to delineate resource conservation: this is the reason for the factorization of the identity. For general lattices, sums are not necessarily sensible. In certain cases, one could imagine something is not quite resource conservation: the framework is flexible enough to adapt.

\begin{figure}
    \centering
    \includegraphics[width=4.5in]{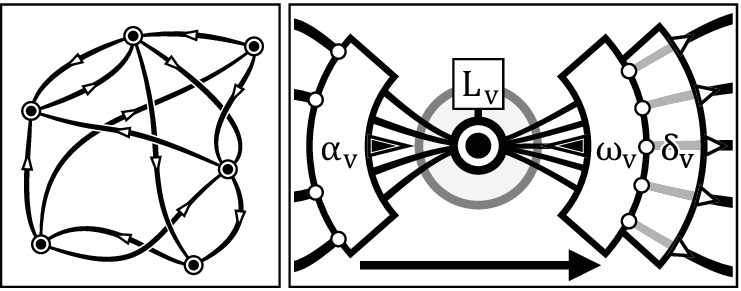}
    \caption{\small A lattice liability network [LLN] begins with a quiver, $Q$ [left]. At each node $v\in V$, there are incoming edges $t^{-1}(v)$ from debtors and outgoing edges $s^{-1}(v)$ to creditors [right]. The LLN $\LLN_Q$ attaches to each vertex $v$ of $Q$ (1) a payment lattice $L_v$, pay-in $\inagg_v$ and pay-out $\outagg_v$ aggregators, and a distributor $\dist_v$ which sends the payment from $v$ to its creditors, respecting nominal liability bounds along edges. Conservation of pay-outs is enforced via $\outagg\circ\dist=\id$. The pay-in aggregator $\inagg_v$ performs any conversions, if needed.}
\label{fig:LLN}
\end{figure}

Note that both the distribution pair $(\dist_v,\outagg_v)$ and the nominal liabilities $\liab_e$ are indexed in the ``currency'' of the paying institution --- that is, they operate within the source lattice $L_{s(e)}$. This natural typing ensures that nodes handle distribution of their resources in their local lattice structure before any conversion occurs during transfer to other nodes.

The factorization condition $\outagg_v\circ\dist_v = \id$ ensures a basic form of resource conservation: what is distributed must aggregate back to the original resources. However, in classical financial networks, one typically has a stronger notion of conservation: the sum of distributed resources cannot exceed available resources. In our general lattice framework, we lack the arithmetic structure needed to express such a sum. One might consider imposing additional conditions --- such as the sub-distributive property $[\dist_v(x\vee y)]_e \leq [\dist_v(x)]_e \vee [\dist_v(y)]_e$ --- to prevent distributors from effectively duplicating resources. While not strictly necessary for our existence results, such conditions align with the economic principle that resources cannot be created by mere redistribution. The challenge lies in formulating the right conservation principle when working with abstract lattices that may lack natural additive structure. 

The generality of this framework may seem imposing, but it naturally specializes to the classical Eisenberg-Noe model when all payment lattices are ordered intervals: see \S\ref{sec:EN-LLN}. The complexity of Definition~\ref{def:LLN} arises precisely because these natural operations on real numbers need explicit analogues when working with more general lattices. Subsequent sections will demonstrate how this abstraction allows us to model everything from multiple-currency obligations to fuzzy payment systems while preserving the fundamental fixed-point structure that made the original Eisenberg-Noe framework so powerful.

\section{Clearing Sections}
\label{sec:clearing}

The core insight of the Eisenberg-Noe model is that a financially consistent set of payments must simultaneously satisfy local constraints at each node while maintaining global consistency across the network. In our more general setting, we seek an analogous notion of network-wide consistency where each node's state respects both its local lattice structure and its relationships with neighboring nodes. This leads us to the concept of a clearing section -- an assignment of lattice elements to vertices that is compatible with all conversion operators, aggregation rules, and limited liabilities. Just as a clearing vector in the classical model represents a feasible set of interbank payments, a clearing section represents a feasible configuration of the entire network that respects the richer structure we have imposed.

\begin{definition}[Payments]
\label{def:edge-payment}
Given a lattice liability network $\LLN_Q$, the \emph{payment lattice} is defined as
\begin{equation}
\label{eq:vertex-payment}
    C^0 = \prod_{v\in V}L_v ,
\end{equation}
with an element $\mathbf{x}=(x_v)\in C^0$ representing a system of institutional payments. Such a payment $\mathbf{x}\in C^0$ can be broken down into \emph{edge payments}. For each $e\in E$ define
\begin{equation}
\label{eq:edge-payment}
    \epymt_e = [\dist_{s(e)}(x_{s(e)})]_e ,
\end{equation}
the amount paid by institution $s(e)$ to institution $t(e)$ along edge $e$ according the payment $\mathbf{x}$. 
\end{definition}

Using aggegrators and distributors to break down payments into a flow of obligated funds allows for a notion of conservation or balance. This, borrowing language from network sheaves, leads to our definition of clearing for an LLN.

\begin{definition}[Clearing Section]
\label{def:clearing-section}
Given a lattice liability network $\LLN_Q$, a \emph{clearing section} is a payment $\mathbf{x}\in C^0$ such that at every vertex $v\in V$:
\begin{equation}
\label{eq:clearing}
    x_v = \inagg_v\left(\left(\epymt_e\right)_{e\in t^{-1}(v)}\right) .
\end{equation}
\end{definition}

That is, the net pay-out $x_v$ at $v$ equals the aggregate pay-in along incoming edges at $v$: a conservation of funds flowing in and out, with the aggregator $\inagg_v$ combining incoming funds.\footnote{See the next section for how exogenous resources and overflows can be adapted.}

\begin{lemma}
\label{lem:nominal}
Clearing sections satisfy the nominal liability bounds.
\end{lemma}
{\em Proof:} Given a clearing section $\mathbf{x}$, and any edge $e$ emanating from vertex $v$, the payment along $e$ satisfies 
\[ 
    \epymt_e = 
    [\dist_v(x_v)]_e \leq 
    [\dist_v(\top_v)]_e \leq 
    \liab_e ,
\]
by Equations (\ref{eq:edge-payment}) and (\ref{eq:Bounded-Liability}) and the monotonicity of distributors. Thus the nominal liability bounds are enforced.
\qed

Clearing sections are figuratively sections -- globally consistent choices of data over vertices. At each vertex, the aggregated incoming resources (after conversion) exactly match the total distributed outgoing resources. The terminology explicitly evokes global sections of network sheaves, where consistency conditions require precise matching of data across edges. Indeed, a clearing section represents a type of equilibrium where no resources are created or destroyed within the network --- every unit that flows out of one vertex must flow into another, modulo the conversion operators. This conservation property distinguishes clearing sections from arbitrary assignments of lattice elements to vertices and reflects the fundamental nature of payment systems: in a consistent state, all transfers must balance. Note that this perfect balance property suggests we will need to augment our framework to handle scenarios with excess resources or losses.

\section{Exogenous Resources}
\label{sec:exo}

A key aspect of financial networks and other resource flow systems is that nodes often have access to resources beyond those received from other nodes in the network. Similarly, nodes may have excess resources not allocated to obligations. Rather than augmenting our base model with additional vertices (which introduces lattice conversion challenges), we incorporate these features directly through self-loops in the quiver structure.

\begin{definition}[Resource-Augmented LLN]
\label{def:resourceLLN}
Given a lattice liability network $\LLN_Q$ with quiver $Q = (V, E, s, t)$, the \emph{resource-augmented network} $\LLN_{Q^+}$ is constructed by adding two self-loops to each vertex $v \in V$:
\begin{enumerate}
\item An \emph{exogenous resource loop} $v \xrightarrow{e_v^{\text{in}}} v$ with nominal liability $\liab_{e_v^{\text{in}}} = \iota_v \in L_v$, where $\iota_v$ represents the exogenous resources available to node $v$.

\item An \emph{overflow resource loop} $v \xrightarrow{e_v^{\text{out}}} v$ with nominal liability $\liab_{e_v^{\text{out}}} = \top_v \in L_v$, where $\top_v$ is the maximal element in $L_v$.
\end{enumerate}

Formally, 
\[
Q^+ = (V, E \cup E^+, s, t)
\]
where $E^+ = \{e_v^{\text{in}}, e_v^{\text{out}} : v \in V\}$ consists of self-loops with $s(e_v^{\text{in}}) = t(e_v^{\text{in}}) = s(e_v^{\text{out}}) = t(e_v^{\text{out}}) = v$.
\end{definition}

To implement this dual self-loop approach, we modify the aggregator and distributor operators as follows:

\begin{enumerate}
\item The distributor $\dist_v$ allocates resources to regular obligations first according to domain-specific rules, then routes any surplus to the overflow loop. The exogenous resource $\iota_v$ is ``created'' {\em ex nihilo} by the distributor and sent to the resource loop. 

\item The pay-out aggregator $\outagg_v$ accommodates the overflow (for conservation purposes) but completely ignores the exogenous resource loop from its aggregation. In this way, the exogenous resource is spontaneously generated.

\item The pay-in aggregator $\inagg_v$ is modified to incorporate exogenous resources while completely ignoring the overflow loop. In this way, overflow resources are ``destroyed'' and unavailable for utilization. 
\end{enumerate}

The specific form of augmented aggregators and distributors depends on the lattice structure. For a classical E-N model with ordered intervals, simple addition operations suffice. More sophisticated lattices may incorporate exogenous resources creatively. So long as monotonicity is preserved by these augmented operators, the LLN will remain functional.

\begin{figure}
    \centering
    \includegraphics[width=4.5in]{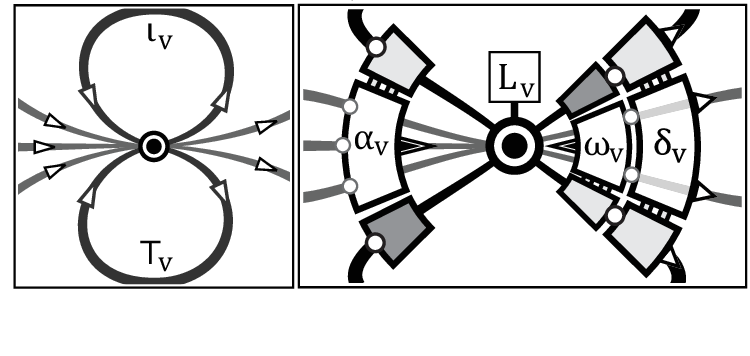}
    \caption{\small [Left] A resource-augmented LLN adds two self-loops to the quiver $Q$ at vertex $v$: a \emph{resource loop} (top) and an \emph{overflow loop} (bottom). [Right] For the resource loop, the distributor $\dist_v$ creates $\iota_v$ and sends it along the loop, but ignores it in the pay-out aggregator $\outagg_v$; the pay-in aggregator $\inagg_v$ can incorporate $\iota_v$ as a (self-)payment. For the overflow loop, the distributor and pay-out aggregator send an overflow amount to the loop; its pay-in aggregator ignores it, creating a sink for the overflow resources.}
\label{fig:resources}
\end{figure}

This approach cleanly separates the introduction of exogenous resources from the handling of overflow resources. The exogenous resource loop acts as a source, injecting $\iota_v$ into the system without being counted in the output aggregation. The overflow resource loop acts as a sink, absorbing any excess resources that remain after meeting obligations.
This ensures that the clearing condition at each vertex accurately reflects both the resources available to it and its capacity to fulfill obligations.

\begin{lemma}[Resource Balanced Clearing]
\label{lem:resource-balance}
Let $\mathbf{x}$ be a clearing section for the resource-augmented lattice liability network $\LLN_{Q^+}$. Then for each vertex $v \in V$, exogenous resources $\iota_v$ are properly incorporated, regular obligations are satisfied according to the distributor's rules, and excess resources are routed to the overflow loop while maintaining $\outagg_v \circ \dist_v = \id_{L_v}$.
\end{lemma}

In what follows, we will use this augmented construction when discussing systems with exogenous resources, often implicitly. Unless otherwise specified, references to a lattice liability network should be understood to include the necessary resource augmentation.

\section{Existence Theorem for Clearing Sections}
\label{sec:MAIN-THM}

Having established the framework of lattice liability networks, we turn to the fundamental question of existence: given such a network, does a clearing section always exist? This question is non-trivial because clearing sections must simultaneously satisfy both local constraints at each node and global consistency conditions across the network. In the classical Eisenberg-Noe model, existence follows from the Tarski fixed point theorem applied to a monotone operator on a product of real intervals. In our more general setting, we must carefully construct an appropriate complete lattice and show that the clearing conditions can be encoded as a suitable monotone operator.

Having established the framework of lattice liability networks, we now address the fundamental question of existence of clearing sections. A clearing section represents a consistent global state of the network where all local constraints are simultaneously satisfied. The following theorem establishes not only existence but also the structure of the set of all clearing sections.

\begin{theorem}[Existence of Clearing Sections]
\label{thm:existence}
Let $\LLN_Q$ be a lattice liability network with quiver $Q = (V,E,s,t)$. Then the set of clearing sections forms a nonempty complete lattice under the product order.
\end{theorem}

{\em Proof:} Consider $C^0 = \prod_{v \in V} L_v$, the payment lattice. Since each $L_v$ is complete, $C^0$ is complete under the product order. Define a payment map $\Phi: C^0 \to C^0$ which updates every payment according to its net in-flux of funds. That is, at each $v\in V$,
\begin{equation}
\label{eq:Phi}
    [\Phi(\mathbf{x})]_v = 
    \inagg_v\left(\left(\epymt_e\right)_{e\in t^{-1}(v)}\right) .
\end{equation}
By definition, the fixed points $\Fix(\Phi)$ are precisely the clearing sections of $\LLN_Q$.

To apply Tarski's Fixed Point Theorem, we need only show that $\Phi$ is monotone. Let $\mathbf{x} \leq \mathbf{y}$ in $C^0$. Then:

1. For each vertex $v$ and edge $e \in s^{-1}(v)$, since $x_v \leq y_v$ and $\dist_v$ is monotone:
\[
\epymt_e = 
[\dist_v(x_v)]_e  \leq 
[\dist_v(y_v)]_e  \leq 
\liab_e
\]

2. Since $\inagg_v$ is monotone in each coordinate:
\[
[\Phi(\mathbf{x})]_v \leq [\Phi(\mathbf{y})]_v
\]
for each vertex $v$. Therefore $\Phi$ is monotone on $C^0$. By Tarski's Fixed Point Theorem, $\Fix(\Phi)$ forms a nonempty complete lattice.
\qed

\begin{corollary}
\label{cor:existence}
Let $\LLN_Q$ be a lattice liability network. Then:
\begin{enumerate}
\item There exists both a least and a greatest clearing section;
\item Any two clearing sections have a uniquely-defined meet and join that are also clearing sections;
\end{enumerate}
\end{corollary}

\section{Distributed Computation of Clearing Sections}
\label{sec:computational}

The existence of clearing sections as guaranteed by Theorem~\ref{thm:existence} naturally leads to questions of computation: how can these clearing sections be found in practice, particularly in large-scale networks? This question becomes especially relevant in systems where entities are unwilling or unable to share complete information about their financial positions, or where no central authority exists to coordinate the clearing process. Many real-world liability networks -- from interbank payment systems to supply chains -- operate through bilateral interactions without global coordination.

We now demonstrate that clearing sections can be computed through a distributed algorithm where each node requires only local information and communication with immediate neighbors. This aligns with practical realities of many network systems while preserving the mathematical guarantees of our framework.\footnote{Computing clearing sections with privacy guarantees in an interesting future direction.}

\begin{algorithm}[H] 
\caption{Distributed Computation of Clearing Sections}
\label{alg:distributed}
\SetKwInOut{Init}{Initialize}
\Init{Each node $v \in V$ sets $x_v^{(0)} \in L_v$}
\For{each iteration $n \geq 0$}{
  \For{each node $v \in V$}{
    Compute $\epymt_e^{(n)} = [\dist_v(x_v^{(n)})]_e$ for all $e \in s^{-1}(v)$\;
    Send $\epymt_e^{(n)}$ to the target node $t(e)$\;
    Receive $\epymt_e^{(n)}$ for all $e \in t^{-1}(v)$\;
    Update $x_v^{(n+1)} = \inagg_v\left(\left(\epymt_e^{(n)}\right)_{e \in t^{-1}(v)}\right)$\;
  }
  \If{$x_v^{(n+1)} = x_v^{(n)}$ for all $v \in V$}{
    \textbf{break}\;
  }
}
\end{algorithm}

\begin{theorem}[Distributed Computation of Clearing Sections]
\label{thm:distributed}
Let $\LLN_Q$ be a lattice liability network with quiver $Q = (V,E,s,t)$. Consider the distributed iterative process described in Algorithm \ref{alg:distributed}. Then:
\begin{enumerate}
\item If $x_v^{(0)} = \bot_v$ for all $v \in V$, the process converges to the least clearing section (corresponding to the least fixed point of $\Phi$).
\item If $x_v^{(0)} = \top_v$ for all $v \in V$, the process converges to the greatest clearing section (corresponding to the greatest fixed point of $\Phi$).
\item For finite lattices, when starting from any initial condition $\mathbf{x}^{(0)} \in C^0$, the process converges to some clearing section.
\end{enumerate}
For finite lattices, convergence occurs in at most $O(\sum_{v \in V} h_v)$ iterations, where $h_v$ is the height of lattice $L_v$. For infinite lattices, convergence from $\bot$ or $\top$ is guaranteed if all aggregators and distributors are Scott-continuous (preserving suprema of directed sets).
\end{theorem}

{\em Proof:} 
The distributed process described in Algorithm \ref{alg:distributed} implements the iteration $\mathbf{x}^{(n+1)} = \Phi(\mathbf{x}^{(n)})$ where $\Phi$ is the payment map defined in the proof of Theorem~\ref{thm:existence}. Each node's update rule:

\[
x_v^{(n+1)} = \inagg_v\left(\left(\epymt_e^{(n)}\right)_{e\in t^{-1}(v)}\right)
\]

where $\epymt_e^{(n)} = [\dist_{s(e)}(x_{s(e)}^{(n)})]_e$ is precisely the component-wise definition of $\Phi$ in Equation~\ref{eq:Phi}.

For statements (1) and (2), the convergence properties follow from the Kleene-Tarski Theorem (see Appendix \ref{sec:chaincomplete}, Theorem \ref{thm:chainTarski}). When starting from $\mathbf{x}^{(0)} = \mathbf{\bot}$, we obtain an increasing sequence $\mathbf{x}^{(0)} \leq \mathbf{x}^{(1)} \leq \mathbf{x}^{(2)} \leq \cdots$ that converges to the least fixed point of $\Phi$. When starting from $\mathbf{x}^{(0)} = \mathbf{\top}$, we obtain a decreasing sequence that converges to the greatest fixed point.

For statement (3) regarding finite lattices, we proceed as follows. Let $L_v$ be a finite lattice for each vertex $v \in V$. Then the product lattice $C^0 = \prod_{v \in V} L_v$ is also finite. For any initial condition $\mathbf{x}^{(0)} \in C^0$, consider the sequence $\mathbf{x}^{(n+1)} = \Phi(\mathbf{x}^{(n)})$.

Since $C^0$ is finite, this sequence must either reach a fixed point or enter a cycle. Suppose, for contradiction, that there exists a cycle of length $k > 1$:
\[
\mathbf{x}^{(n)} \mapsto \mathbf{x}^{(n+1)} \mapsto \cdots \mapsto \mathbf{x}^{(n+k-1)} \mapsto \mathbf{x}^{(n)} 
\]

Let $\mathbf{z} = \mathbf{x}^{(n)} \vee \mathbf{x}^{(n+1)} \vee \cdots \vee \mathbf{x}^{(n+k-1)}$ be the join of all elements in this cycle. By monotonicity of $\Phi$, for any $i \in \{0,1,...,k-1\}$:
\[
\Phi(\mathbf{x}^{(n+i)}) \leq \Phi(\mathbf{z})
\]

But $\Phi(\mathbf{x}^{(n+i)}) = \mathbf{x}^{(n+i+1)}$ (where indices are taken modulo $k$), so each element in the cycle is $\leq \Phi(\mathbf{z})$. Therefore:
\[
\mathbf{z} = \mathbf{x}^{(n)} \vee \mathbf{x}^{(n+1)} \vee \cdots \vee \mathbf{x}^{(n+k-1)} \leq \Phi(\mathbf{z})
\]

Now, consider the sequence $\mathbf{z}, \Phi(\mathbf{z}), \Phi^2(\mathbf{z}),...$. Since $\mathbf{z} \leq \Phi(\mathbf{z})$ and $\Phi$ is monotone, this sequence is increasing. As $C^0$ is finite, this sequence must reach a fixed point $\mathbf{w} = \Phi(\mathbf{w})$ after finitely many iterations.

For each $i$, we have $\mathbf{x}^{(n+i)} \leq \mathbf{z} \leq \mathbf{w}$. By monotonicity, $\Phi(\mathbf{x}^{(n+i)}) \leq \Phi(\mathbf{w}) = \mathbf{w}$, thus $\mathbf{x}^{(n+i+1)} \leq \mathbf{w}$ for all $i$. Therefore, the entire cycle is bounded above by the fixed point $\mathbf{w}$.

Let $j$ be such that $\mathbf{x}^{(n+j)}$ is maximal in the cycle (such a $j$ exists because the cycle has finitely many elements). Then:
\[
\mathbf{x}^{(n+j)} \leq \Phi(\mathbf{x}^{(n+j)}) = \mathbf{x}^{(n+j+1)}
\]

But since $\mathbf{x}^{(n+j)}$ is maximal in the cycle, we must have $\mathbf{x}^{(n+j)} = \mathbf{x}^{(n+j+1)}$, which means $\mathbf{x}^{(n+j)}$ is a fixed point. This contradicts our assumption that the elements form a proper cycle of length $k > 1$.

Thus, any sequence must reach a fixed point after finitely many iterations, regardless of the initial condition.

For the complexity bound, in each iteration that does not reach a fixed point, at least one component must change. Since each lattice $L_v$ has finite height $h_v$ and each component can change at most $h_v$ times in a monotonic sequence, the maximum number of iterations before reaching a fixed point is bounded by $\sum_{v \in V} h_v$.

For infinite lattices, the guarantee of convergence from arbitrary starting points does not generally hold without additional assumptions. However, when starting from $\bot$ or $\top$, Scott-continuity ensures that $\Phi$ preserves suprema of directed sets (including chains), guaranteeing convergence to the least or greatest fixed point, respectively.
\qed

The ability to compute clearing sections through purely local interactions has important practical implications. The distributed algorithm described above offers several key advantages:

\begin{remark}[Practical Implementation]
The distributed algorithm offers three notable benefits for real-world applications. First, it preserves privacy, as nodes share only payment amounts with direct counterparties rather than revealing internal states or decision rules. Second, it minimizes coordination requirements, as nodes need only synchronize iterations with immediate neighbors. Third, it provides flexibility in targeting different clearing sections based on initialization strategy.

Implementation in large networks presents specific challenges. For termination detection without a central coordinator, nodes can employ a signaling protocol where each node indicates when its state has stabilized. Global convergence is then detected when all such signals have propagated through the network, similar to Dijkstra's ``diffusing computations'' \cite{dijkstra1980termination}. For asynchronous systems where nodes update at different rates, the algorithm can be adapted to use the most recent available information from neighbors, though the specific fixed point reached may depend on the update sequence.

For infinite lattices with continuous domains, practical implementations typically replace exact equality testing with approximate termination conditions based on appropriate distance measures between consecutive iterations.
\end{remark}

\begin{remark}[Computational Complexity]
The computational properties of the algorithm depend directly on the lattice structure. For finite lattices, each iteration requires $O(|E|)$ operations across all nodes. With convergence guaranteed in at most $O(\sum_{v \in V} h_v)$ iterations, the overall worst-case time complexity is $O(|E| \cdot \sum_{v \in V} h_v)$.

For infinite lattices, when starting from extremal points ($\bot$ or $\top$) with Scott-continuous operators, convergence to the corresponding extremal fixed point is guaranteed, but may require unbounded iterations without additional structural assumptions.
\end{remark}

This combination of theoretical guarantees and practical adaptability makes our framework particularly suitable for decentralized and privacy-sensitive applications where global coordination is impractical or undesirable.

\section{Classical Eisenberg-Noe}
\label{sec:EN-LLN}

The classical Eisenberg-Noe model emerges naturally from our framework as a simple special case. Consider a financial system with $n$ banks. We construct a lattice liability network $\LLN_Q$ with quiver $Q = (V,E,s,t)$ where:

\begin{enumerate}
\item The vertex set $V = \{1,\ldots,n\}$ represents banks;
\item The edge set $E$ consists of ordered pairs $(i,j)$ where bank $i$ has an obligation to bank $j$.
\end{enumerate}

Following Section~\ref{sec:exo}, we augment this to $Q^+ = (V, E \cup E^+, s, t)$ with $E^+ = \{e_v^{\text{in}}, e_v^{\text{out}} : v \in V\}$, where $e_v^{\text{in}}$ and $e_v^{\text{out}}$ are self-loops representing exogenous resources and overflow, respectively.

For each vertex $v \in V$, we assign the complete lattice $L_v = [0,\infty]$ with the usual ordering. The nominal liabilities are:
\begin{itemize}
\item For each interbank edge $e \in E$: $\liab_e$ equals the nominal obligation from $s(e)$ to $t(e)$;
\item For each exogenous resource loop $e_v^{\text{in}} \in E^+$: $\liab_{e_v^{\text{in}}} = \iota_v$ equals bank $v$'s external assets;
\item For each overflow loop $e_v^{\text{out}} \in E^+$: $\liab_{e_v^{\text{out}}} = \infty$ (effectively unbounded).
\end{itemize}

For any payment vector $\mathbf{x}=(x_v)$, the pay-in aggregator $\inagg_v$ sums all incoming payments except payments from the overflow loop:
\[
    \inagg_v\left(\left(\epymt_e\right)_{e \in t^{-1}(v)}\right) 
    = 
    \epymt_{e_v^{\text{in}}} + \sum_{e \in t^{-1}(v) \setminus \{e_v^{\text{in}}, e_v^{\text{out}}\}} \epymt_e 
    =
    \iota_v + \sum_{e \in t^{-1}(v) \cap E} \epymt_e.
\]

The distributor $\dist_v$ implements proportional sharing for interbank liabilities while routing any excess funds to the overflow loop. Let $\bar{\liab}_v = \sum_{e \in s^{-1}(v) \cap E} \liab_e$ be the total interbank liability, and $\pi_e = \liab_e/\bar{\liab}_v$ be the proportional factor for interbank edge $e \in s^{-1}(v) \cap E$. Then,
\[
    [\dist_v(x_v)]_e = 
    \begin{cases}
        \pi_e \cdot \min\{x_v, \bar{\liab}_v\} & \text{if } e \in s^{-1}(v) \cap E \\
        \iota_v & \text{if } e = e_v^{\text{in}} \\
        \max\{0, x_v - \bar{\liab}_v\} & \text{if } e = e_v^{\text{out}}
    \end{cases}.
\]

The pay-out aggregator $\outagg_v$ sums the payments from interbank edges and the overflow loop (but ignores the exogenous resource loop):
\[
\outagg_v\left(\left(\epymt_e\right)_{e \in s^{-1}(v)}\right) 
= 
\sum_{e \in s^{-1}(v) \cap E} \epymt_e + \epymt_{e_v^{\text{out}}}.
\]

It is straightforward to verify that $\outagg_v \circ \dist_v = \id_{L_v}$, satisfying our resource conservation requirement.

A clearing section $\mathbf{x}$ on this network then satisfies, for each bank $v \in V$
\[
x_v = \iota_v + \sum_{e \in t^{-1}(v) \cap E} \epymt_e
\]
where each $\epymt_e$ is defined using the distributor function. Substituting the definition of $\epymt_e$, we get
\[
x_v = 
\iota_v + \sum_{e \in t^{-1}(v) \cap E} \pi_e \cdot \min\{x_{s(e)}, \bar{\liab}_{s(e)}\}.
\]

Since $\min\{x_v, \bar{\liab}_v\} = \min\{\iota_v + \sum_{e \in t^{-1}(v) \cap E} \epymt_e, \bar{\liab}_v\}$, this is equivalent to
\[
x_v = 
\min\!\left\{
\bar{\liab}_v
\; , \; 
\iota_v + \sum_{e \in t^{-1}(v) \cap E} \pi_e \cdot \min\{x_{s(e)}, \bar{\liab}_{s(e)}\}
\right\}
\]

which is exactly the classical Eisenberg-Noe clearing condition (Equation \ref{eq:EN}).

\begin{corollary}
The clearing sections of the above lattice liability network correspond precisely to the clearing payment vectors of the classical Eisenberg-Noe model.
\end{corollary}

\section{Multiple Currencies}
\label{sec:currencies}

The framework of lattice liability networks elegantly handles international financial systems where each entity operates primarily in its native currency. Consider a network of sovereign nations where each node $v \in V$ has an associated currency (e.g., USD, EUR, JPY). The payment lattice for node $v$ is simply $L_v = [0,\infty]$, representing amounts in $v$'s native currency. Following Section~\ref{sec:exo}, we augment each node with exogenous resource and overflow loops, allowing nations to incorporate external assets and handle excess payments. This self-loop mechanism naturally captures the fact that a nation's reserves and surpluses are typically denominated in its own currency.

The nominal liabilities $\liab_e$ for an edge $e: v \to w$ specify an amount in the source node's currency. For instance, if nation $v$ owes nation $w$ an amount $\liab_e$, this obligation is denominated in $v$'s currency. The currency conversion happens at the receiving end through the pay-in aggregator.

The pay-in aggregator $\inagg_w$ at node $w$ must convert all incoming payments to $w$'s native currency before combining them. For each incoming edge $e: v \to w$, let $\conv_e: L_v \to L_w$ be a currency conversion operator that transforms payments from $v$'s currency to $w$'s currency (a monotone lattice map). Then:
\[
    \inagg_w\left(\left(\epymt_e\right)_{e \in t^{-1}(w)}\right) 
    = 
    \sum_{e \in t^{-1}(w)} \conv_e(\epymt_e)
\]

The conversion operators $\conv_e$ can model market realities such as exchange rate spreads and volume-dependent slippage. For example, a simple linear conversion would be $\conv_e(x) = \rho_e x$ where $\rho_e$ is the exchange rate from $v$'s currency to $w$'s currency. More realistic nonlinear conversions might take the form $\conv_e(x) = \rho_e x(1 - s(x))$ where $s(x)$ is an increasing function representing slippage. Such functions are clearly monotone (more input currency produces more output currency).

Each node's distributor $\dist_v$ operates entirely in the node's native currency, allocating available resources to meet obligations. Since all of $v$'s obligations are denominated in its own currency, no further conversion is needed at the distribution stage. The corresponding pay-out aggregator $\outagg_v$ similarly works purely in $v$'s native currency.

A clearing section in this context represents a consistent pattern of international payments where each nation pays in its domestic currency and recipients handle the conversion. This matches reality: when Japan pays a USD-denominated debt, the yen must be converted to dollars, typically at the receiving end.

This formulation raises interesting theoretical questions about the relationship between exchange rate dynamics and clearing section structure. When conversion operators are linear, the system may inherit uniqueness properties similar to the classical Eisenberg-Noe model. However, nonlinear effects from slippage or market impact could create multiple equilibria representing different patterns of currency flows, potentially modeling currency crises where sudden shifts in exchange rates trigger cascading defaults.

\section{Automated Market Makers in Decentralized Finance}
\label{sec:amm}

Decentralized finance (DeFi) offers a rich domain for applying lattice liability networks, particularly in the context of Automated Market Makers (AMMs) - the cornerstone mechanism enabling decentralized token exchange. AMMs operate as constant function market makers where liquidity providers deposit token pairs into pools, and traders exchange tokens against these pools according to deterministic pricing functions. We demonstrate how the mathematical structure of interconnected AMM pools naturally fits our lattice liability framework.

Consider a network of AMM pools represented as a quiver $Q = (V, E, s, t)$ where vertices $v \in V$ represent individual liquidity pools (e.g., ETH-USDC, ETH-DAI, DAI-USDC), and edges $e \in E$ represent potential token flows between pools.

For each liquidity pool $v \in V$, we define the payment lattice as: $L_v = [0, \infty]^2$,
where each element $(x,y) \in L_v$ represents the reserves of the two tokens in the pool, and the product order is maintained: greater reserves provide better liquidity depth and reduce price impact for traders.

For each edge $e: v \to w$ connecting two pools that share a common token, the nominal liability $\liab_e \in L_{s(e)}$ represents the maximum allowable token flow between the pools, bounded by practical or protocol-imposed limits. For instance, if pool $v$ contains tokens A and B, while pool $w$ contains tokens B and C, the nominal liability might be $\liab_e = (0, M)$, where $M$ is the maximum amount of token B that can flow from pool $v$ to pool $w$ in a single transaction (possibly determined by gas limits or slippage parameters).\footnote{In practice, more complex scenarios can leverage the currency conversion mechanisms described in Section~\ref{sec:currencies}. For instance, if pools $v$ and $w$ share no common tokens, a trade from $v$ to $w$ would involve intermediate conversions, perhaps across a routing path.}

Following Section~\ref{sec:exo}, we can augment each vertex with exogenous resource and overflow loops to model external liquidity provision and removal. These self-loops capture how traders and liquidity providers interact with the pools from outside the network.

The key to modeling AMM behavior lies in the pay-in aggregator and distributor-aggregator pair, which must respect a product constraint: each pool operates according to the constant product formula $x \cdot y = \kappa$, where $x$ and $y$ are token reserves and $\kappa$ is a constant.
For a pool $v$ with current reserves $(x_v, y_v)$, the pay-in aggregator combines incoming token flows:

\[
\inagg_v\left(\left(\epymt_e\right)_{e \in t^{-1}(v)}\right) = \left(x_v + \sum_{e \in t^{-1}(v)} f_e(\epymt_e), y_v + \sum_{e \in t^{-1}(v)} g_e(\epymt_e)\right)
\]

where $f_e$ and $g_e$ are token extraction functions that determine how much of each token type from $\epymt_e$ is relevant to pool $v$. For edges representing direct swaps, typically only one token flows in, affecting only one component.

The distributor $\dist_v: L_v \to L_v^{|s^{-1}(v)|}$ must respect both conservation of tokens and the constant product formula. When a swap occurs from pool $v$ to pool $w$, the distributor ensures that:

\[
[\dist_v(x_v, y_v)]_e = \min\left\{(h_{e,x}, h_{e,y}), \liab_e\right\}
\]

where $(h_{e,x}, h_{e,y})$ represents the token outflow that maintains the constant product $\kappa_v$. For example, if a trader provides $\Delta x$ units of token A to remove token B, then:
\[
h_{e,y} = y_v - \frac{\kappa_v}{x_v + \Delta x}
\]
where $\kappa_v = x_v \cdot y_v$ is the constant product before the trade.

The pay-out aggregator $\outagg_v: L_v^{|s^{-1}(v)|} \to L_v$ then combines these distributed resources:

\[
\outagg_v\left(\left(\epymt_e\right)_{e \in s^{-1}(v)}\right) = \left(x_v - \sum_{e \in s^{-1}(v)} [\epymt_e]_x, y_v - \sum_{e \in s^{-1}(v)} [\epymt_e]_y\right)
\]

The identity condition $\outagg_v \circ \dist_v = \id_{L_v}$ is satisfied naturally since the distributor allocates exactly the outgoing tokens from the pool's reserves, and the aggregator simply accounts for these allocations.

A clearing section $\mathbf{x} = (x_v, y_v)_{v \in V}$ in this context represents a stable configuration of reserves across all pools that is consistent with all token flows. This directly corresponds to the equilibrium state of an interconnected AMM network after a series of trades has executed.

The lattice structure of clearing sections provides insights into several key phenomena in DeFi:

\begin{enumerate}
\item \emph{Path Dependence}: Different sequences of trades between the same pools can lead to different final states, represented by distinct clearing sections.

\item \emph{Arbitrage Opportunities}: If two clearing sections $\mathbf{x}, \mathbf{y} \in \Fix(\Phi)$ satisfy $\mathbf{x} < \mathbf{y}$ (component-wise), their difference represents a potential arbitrage opportunity - a sequence of trades that could move the system from state $\mathbf{x}$ to state $\mathbf{y}$ while extracting profit.

\item \emph{Liquidity Fragmentation}: The greatest clearing section typically represents the most efficient distribution of liquidity across pools, maximizing total reserves while maintaining all constraints.
\end{enumerate}

This formulation naturally captures the concept of Maximal Extractable Value (MEV) - the maximum value that can be extracted by reordering, inserting, or censoring transactions. In lattice terms, MEV represents a distance between clearing sections, particularly between the current state and the greatest clearing section that could be reached through an optimal sequence of trades.

The framework also extends to more complex AMM designs, such as concentrated liquidity pools where the payment lattice would represent liquidity distribution across price ranges, or hybrid pools with dynamically adjusting curves. In each case, the lattice structure provides a unified way to analyze equilibrium states and optimization opportunities in these interconnected, decentralized markets.

\section{Manufacturing and Supply Chains}
\label{sec:supply}

While our earlier examination of currency networks involved exchange operations that preserve the essential nature of value, supply chains fundamentally transform resources into different forms. This transformational aspect makes supply chains a natural yet distinct application of our lattice liability network framework.

We model a supply chain network with a quiver $Q = (V, E, s, t)$ where vertices represent production facilities, warehouses, or distribution centers, and edges represent transportation channels or contractual relationships. Unlike currency networks where each node operates within its native currency, supply chain nodes handle multiple resource types simultaneously (raw materials, intermediate components, finished products), making product lattices an ideal state space representation.

For each node $v \in V$, we define the payment lattice as the product $L_v = \prod_{i=1}^{n_v} [0, c_i^v]$, where $c_i^v$ represents the capacity for resource type $i$ at node $v$. The nominal liability along an edge $e \in E$ from supplier $s(e)$ to customer $t(e)$ represents the contractually obligated delivery $\liab_e$ of resources.

The crux of this example lies in the pay-in aggregator. For each node $v \in V$, the pay-in aggregator $\inagg_v: \prod_{e \in t^{-1}(v)} L_{s(e)} \to L_v$ first combines incoming resources from all incoming edges and then applies a manufacturing transformation:

\[
\inagg_v\left(\left(\epymt_e\right)_{e \in t^{-1}(v)}\right) = \trans_v\left(\left(\epymt_e\right)_{e \in t^{-1}(v)}\right)
\]

where $\trans_v: \prod_{e \in t^{-1}(v)} L_{s(e)} \to L_v$ is a manufacturing transformation function that converts input resources into outputs. This operation is not required to conserve resources in the way financial systems do -- raw materials might be consumed to produce fewer finished goods by weight or volume. Only monotonicity is required: increasing manufacturing inputs (in any component) must not decrease production outputs, though the increase need not be strict.

The distributor $\dist_v$ at node $v$ allocates produced resources to downstream customers according to contractual obligations and priorities:
\[
[\dist_v(x_v)]_e = \min\{\pi_e \cdot x_v, \liab_e\}
\]
where $\pi_e$ represents distribution priorities or proportions for each outgoing edge.

The corresponding pay-out aggregator $\outagg_v$ combines distributed resources:
\[
\outagg_v\left(\left(\epymt_e\right)_{e \in s^{-1}(v)}\right) = \sum_{e \in s^{-1}(v)} \epymt_e
\]
This satisfies the factorization identity $\outagg_v \circ \dist_v = \id_{L_v}$ required by our framework.

A clearing section $\mathbf{x} = (x_v)_{v \in V}$ in this context represents a feasible steady-state flow of resources through the supply chain network. If $\mathbf{x}, \mathbf{y} \in \Fix(\Phi)$ with $\mathbf{x} < \mathbf{y}$, then the gap between $\mathbf{x}$ and $\mathbf{y}$ represents untapped capacity that could be activated through appropriate coordination.

In this simple setup, the payment lattice $C^0$ is one large product of ordered intervals. The framework can accommodate more complex lattice structures that represent sophisticated manufacturing constraints, such as partial orders reflecting precedence constraints in assembly processes, or lattices encoding feasible regions for chemical processes with complex reaction dynamics. These extensions would enable modeling advanced manufacturing systems where the state space cannot be decomposed into independent components, yet still maintains the crucial lattice structure needed for our theoretical guarantees.

To incorporate external resource supplies, inventory storage, waste products, or recycling streams, one could augment this basic model following the approach in \S\ref{sec:exo}. This would introduce self-loops that model resource inputs and outputs not connected to other nodes in the network, while preserving the mathematical elegance of the fixed-point formulation.

\section{Residuated Lattices \& Logics}
\label{sec:fuzzy}

The classical Eisenberg-Noe model treats payments as purely quantitative values. However, real-world obligations often involve qualitative dimensions with logical relationships that affect their fulfillment and valuation. In financial systems, payments may vary not only in amount but also in attributes such as timeliness, completeness, or compliance with contractual terms, each of which may require logical operations to process. We demonstrate how our framework naturally accommodates such settings using residuated lattice factors that have monoidal $\otimes$ and adjoint $\rightarrow$ operations generalizing conjunction and implication: see Appendix~\ref{sec:residuation} for background. Such lattice coefficients provide greater expressive power while preserving mathematical tractability.

\begin{definition}[Residuated Payment Lattice]
Let $S_v$ be a finite set of payment attributes for vertex $v$ (e.g., timeliness, completeness, method of payment), and let $(\mathcal{R}, \wedge, \vee, \otimes, \rightarrow, 0, 1)$ be a complete residuated lattice as defined in Appendix~\ref{sec:residuation}. The payment lattice at $v$ is defined as:
\[
L_v = [0,\infty] \times \mathcal{R}^{S_v}
\]
where $\mathcal{R}^{S_v}$ denotes the set of functions from $S_v$ to $\mathcal{R}$, ordered pointwise: $\mu \leq \mu'$ if and only if $\mu(s) \leq \mu'(s)$ for all $s \in S_v$.
\end{definition}

This product lattice is complete since both $[0,\infty]$ and $\mathcal{R}^{S_v}$ are complete. The first component represents the payment amount, while the second component encodes the degree to which various quality attributes are satisfied according to the logical structure of $\mathcal{R}$. While the product lattice is not residuated, we can use addition on the first factor and the residuated structure on the remaining factors. 

The choice of residuated lattice $\mathcal{R}$ determines how attribute values combine and interact. As detailed in Appendix~\ref{sec:residuation}, common choices include G\"odel logic with $\otimes = \min$, \L{}ukasiewicz logic with bounded sum, or Product logic with standard multiplication, each capturing different aspects of fuzzy reasoning appropriate for different application domains.

\begin{definition}[Residuated Nominal Liabilities]
For each edge $e \in E$, the nominal liability $\liab_e = (q_e, \mu_e)$ specifies both a required payment amount $q_e \in [0,\infty]$ and required quality levels $\mu_e \in \mathcal{R}^{S_{s(e)}}$.
\end{definition}

The pay-in and pay-out operations exploit the rich structure of residuated lattices:

\begin{definition}[Residuated Pay-in Aggregator]
For vertex $v$, given incoming payments $(q_e,\mu_e)$ for $e \in t^{-1}(v)$, the pay-in aggregator $\inagg_v: \prod_{e \in t^{-1}(v)} L_{s(e)} \to L_v$ is defined as:
\[
\inagg_v\left(\{(q_e,\mu_e)\}_{e \in t^{-1}(v)}\right) = \left(\sum_{e \in t^{-1}(v)} q_e,\; \bigotimes_{e \in t^{-1}(v)} \conv_e(\mu_e)\right)
\]
where $\conv_e: \mathcal{R}^{S_{s(e)}} \to \mathcal{R}^{S_v}$ is a quality conversion operator that is monotone, and $\bigotimes$ represents the pointwise application of the $\otimes$ operation.
\end{definition}

The pay-in aggregator sums payment quantities while using the $\otimes$ operation to combine quality profiles. This captures the logical conjunction of quality attributes: a payment system must satisfy multiple criteria simultaneously. The $\otimes$ operation from residuated lattices offers nuanced modeling of how partial satisfactions combine.

\begin{definition}[Residuated Distributor-Aggregator Pair]
For vertex $v$, let the proportional payment factor for edge $e \in s^{-1}(v)$ be denoted $\pi_e = q_e / \sum_{e' \in s^{-1}(v)} q_{e'}$. The distributor $\dist_v: L_v \to L_v^{|s^{-1}(v)|+1}$ is defined as:
\[
[\dist_v(q_v,\mu_v)]_e = 
\begin{cases}
\left(\min\{q_v \cdot \pi_e, q_e\}, \mu_v\right) & \text{if } e \in s^{-1}(v) \\
\left(q_v - \sum_{e' \in s^{-1}(v)} \min\{q_v \cdot \pi_e, q_e\}, \mu_v\right) & \text{if } e = e_v^{\text{out}} \text{ (overflow loop)}
\end{cases}
\]

The corresponding pay-out aggregator $\outagg_v: L_v^{|s^{-1}(v)|+1} \to L_v$ is defined as:
\[
\outagg_v\left(\{(q_e,\mu_e)\}_{e \in s^{-1}(v) \cup \{e_v^{\text{out}}\}}\right) = \left(\sum_{e \in s^{-1}(v) \cup \{e_v^{\text{out}}\}} q_e,\; \gamma_v\left(\{\mu_e\}_{e \in s^{-1}(v) \cup \{e_v^{\text{out}}\}}\right)\right)
\]
where $\gamma_v$ is a monotone function that preserves consensus values:
\begin{align*}
\gamma_v\left(\{\mu_e\}_{e \in s^{-1}(v)}\right) = 
\begin{cases}
\mu & \text{if $\mu_e = \mu$ for all $e \in s^{-1}(v)$} \\
\bigotimes_{e \in s^{-1}(v)} \mu_e & \text{otherwise}
\end{cases}
\end{align*}
\end{definition}

The distributor implements proportional payment allocation for the quantity component while copying the quality profile across all outputs. This structure reflects the dual nature of our payment model: quantities are conserved resources that must be allocated proportionally, while quality attributes represent informational characteristics that can be replicated across all payments without diminishment. The pay-out aggregator sums the quantities and applies a special aggregation function $\gamma_v$ to quality attributes. This function preserves the input quality when all quality profiles match (as happens after distribution), while defaulting to the residuated lattice operation $\otimes$ for combining genuinely different quality profiles.

\label{lem:residuated-LLN}
\begin{lemma}
The residuated payment operations satisfy the required properties of a lattice liability network:
\begin{enumerate}
\item The pay-in aggregator $\inagg_v$ is monotone in each coordinate
\item The distributor $\dist_v$ is monotone and respects liability bounds
\item The factorization identity $\outagg_v \circ \dist_v = \id_{L_v}$ holds
\end{enumerate}
\end{lemma}

{\em Proof:} For monotonicity of $\inagg_v$, observe that if $(q_e,\mu_e) \leq (q'_e,\mu'_e)$ for each $e \in t^{-1}(v)$, then $\sum_e q_e \leq \sum_e q'_e$ and $\bigotimes_e \conv_e(\mu_e) \leq \bigotimes_e \conv_e(\mu'_e)$ since $\otimes$ is monotone in both arguments (by Proposition 1 in Appendix~\ref{sec:residuation}) and each $\conv_e$ is monotone. Thus, $\inagg_v$ is monotone in each coordinate.

For the distributor, if $(q_v,\mu_v) \leq (q'_v,\mu'_v)$, then $q_v \cdot \pi_e \leq q'_v \cdot \pi_e$ and $\mu_v \leq \mu'_v$ for each $e \in s^{-1}(v)$, establishing monotonicity. By definition, $[\dist_v(q_v,\mu_v)]_e \leq (q_e, \top) \leq \liab_e$, so liability bounds are respected.

For the factorization identity, we verify that $\outagg_v \circ \dist_v = \id_{L_v}$. Our distributor copies the quality profile $\mu_v$ to each outgoing edge and the overflow loop, producing $\{(q_v \cdot \pi_e, \mu_v)\}_{e \in s^{-1}(v)}$ and $(q_v - \sum_{e \in s^{-1}(v)} \min\{q_v \cdot \pi_e, q_e\}, \mu_v)$ for the overflow loop. When $\outagg_v$ is applied to this output:
\[
\begin{aligned}
\outagg_v\left(\dist_v(q_v,\mu_v)\right) &= \left(\sum_{e \in s^{-1}(v)} \min\{q_v \cdot \pi_e, q_e\} + \left(q_v - \sum_{e \in s^{-1}(v)} \min\{q_v \cdot \pi_e, q_e\}\right), \gamma_v\left(\{\mu_v, \mu_v, \ldots\}\right)\right) \\
&= (q_v, \mu_v)
\end{aligned}
\]

Since all quality profiles are identical ($\mu_e = \mu_v$), the first case of $\gamma_v$ applies, giving $\gamma_v(\{\mu_v, \mu_v, \ldots\}) = \mu_v$. For the quantity component, we have $\sum_{e \in s^{-1}(v)} q_v \cdot \pi_e = q_v \sum_{e \in s^{-1}(v)} \pi_e = q_v$ since $\sum_{e \in s^{-1}(v)} \pi_e = 1$. Therefore:
\[
\outagg_v\left(\dist_v(q_v,\mu_v)\right) = (q_v, \mu_v)
\]

Thus, $\outagg_v \circ \dist_v = \id_{L_v}$ as required.
\qed

A clearing section $\mathbf{x} = \{(q_v, \mu_v)\}_{v \in V}$ in this context represents a network-wide payment configuration that is consistent in both quantity and quality. By Theorem \ref{thm:existence}, such clearing sections exist and form a complete lattice. The greatest clearing section represents the optimal fulfillment of both quantity and quality across the network.

The residuated payment model offers several advantages over conventional approaches:

\begin{enumerate}
\item It naturally handles \emph{partial fulfillment} of quality requirements through the residuated lattice structure, enabling more realistic modeling of contractual obligations
\item It distinguishes between \emph{logical aggregation} (via $\otimes$) and \emph{arithmetic aggregation} (via quantity summation), better reflecting how quality assessments combine in practice
\item It supports different notions of \emph{logical implication} between requirements through the residuum $\rightarrow$, allowing flexible modeling of compatibility and substitution relationships
\item It maintains a clear \emph{separation of concerns} between quantitative payments and qualitative attributes while providing a unified mathematical framework
\end{enumerate}

\section{Residuated Logic in Multi-Attribute Networks}
\label{sec:applications}

The residuated lattice framework introduced in Section~\ref{sec:fuzzy} provides a powerful foundation for modeling complex networks where both quantitative flows and qualitative attributes matter. We explore two concrete applications that demonstrate how our framework naturally accommodates real-world systems where resource flows must satisfy multiple criteria simultaneously.

\subsection{Financial Compliance Network}
\label{subsec:financial-compliance}

\begin{figure}
    \centering
    \includegraphics[width=4.5in]{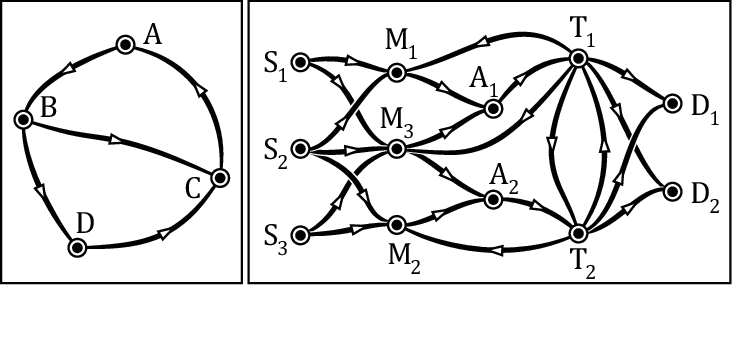}
    \caption{\small A four-bank lattice liability network is based on a quiver with multiple interacting cycles [left]. The payment lattices are all the same: amounts (in $[0,\infty]$) along with a pair of compliance attributes under the \L{}ukasiewicz operations, with nominal liabilities $\liab$ reflecting amount owed and maximal optimistic compliance qualities. A more complex 12-node quiver encodes suppliers ($S$), manufacturers ($M$), assemblers ($A$), testers ($T$), and distributors ($D$) [right]. The payment lattices are again homogeneous: quantities (in $[0,\infty]$) along with three quality attributes under the G\"odel operations, with nominal liabilities $\liab$ reflecting amount owed and maximal optimistic qualities.} 
\label{fig:QUIVERS}
\end{figure}

Consider a four-bank international payment network with banks $A$, $B$, $C$, and $D$ connected by payment obligations. The network contains a cycle $(A,B,C,A)$, with bank $D$ connected to both $B$ and $C$ via cross-cutting edges, creating a more complex topology than a simple chain: see Figure \ref{fig:QUIVERS}[left].

Each bank's state is modeled using a payment lattice $L_v = [0, \infty] \times [0,1]^{\mathcal{A}}$, where $[0, \infty]$ represents payment amounts and $\mathcal{A} = \{a_1, a_2\}$ represents compliance attributes, with $a_1$ for anti-money laundering and $a_2$ for documentation requirements. The \L{}ukasiewicz operations govern how these compliance scores combine across the network:
\[
x \otimes y = \max(0, x + y - 1) 
\quad : \quad
x \rightarrow y = \min(1, 1 - x + y)
\]

This choice captures how compliance quality degrades when information passes through multiple institutions: even if each institution maintains high internal standards, the transmission process itself introduces information loss and administrative friction. Each institution in the chain can at best maintain -- but typically reduces -- the compliance quality it receives.

The nominal liabilities in this network are:
\begin{align*}
\liab_{(A,B)} &= (1000, (0.9, 0.8)) \\
\liab_{(B,C)} &= (800, (0.85, 0.75)) \\
\liab_{(C,A)} &= (900, (0.8, 0.7)) \\
\liab_{(B,D)} &= (500, (0.85, 0.8)) \\
\liab_{(D,C)} &= (400, (0.75, 0.7))
\end{align*}

For each edge, the first component represents the maximum monetary payment, while the pair of compliance factors represents the maximum achievable compliance quality that can be transmitted. Consistent with our lattice framework, higher values represent better states -- higher payment capacity and better compliance (where 1.0 means perfect compliance). The nominal liability thus represents an upper bound on both dimensions.

Consider the calculation at Bank $C$, which, in the best case scenario, receives payments from both $B$ and $D$ equalling:
\begin{align*}
\inagg_C\left( (800, (0.7, 0.7)), (400, (0.75, 0.7)) \right) &= \left( 1200, \left( (0.7 \otimes 0.75), (0.7 \otimes 0.7) \right) \right) \\
&= (1200, (0.45, 0.4))
\end{align*}

Here, $0.7 \otimes 0.75 = \max(0, 0.7 + 0.75 - 1) = 0.45$, and $0.7 \otimes 0.7 = \max(0, 0.7 + 0.7 - 1) = 0.4$.

The dramatic drop in optimal compliance scores (from $(0.7, 0.7)$ and $(0.75, 0.7)$ to $(0.45, 0.4)$) demonstrates the amplification of compliance risk. When computing the complete clearing section for this network, one notes that regardless of how high the initial compliance scores are, the cyclic structure $(A,B,C,A)$ causes all compliance scores to eventually degrade to zero at equilibrium. Each iteration through the cycle applies the \L{}ukasiewicz conjunction again, progressively reducing optimal compliance scores with convergence  to zero.

This reveals the systemic risk in compliance scores: without external intervention in the form of regulatory minimums or periodic compliance refreshes, cyclical networks will inevitably experience complete compliance collapse. The clearing section framework thus highlights why actual regulatory systems must include mechanisms that break this cycle of decay, such as minimum thresholds and independent audits. This failure mode cannot be detected by examining individual institutions in isolation, demonstrating the value of network-level analysis for regulatory design.

\subsection{Supply Chain Quality Network}
\label{subsec:supply-chain}

Consider a manufacturing ecosystem with multiple interlinked production facilities, testing centers, and feedback channels. Unlike the simplified linear chains often used in supply chain models, this network features cross-cutting dependencies and cyclical quality information flows that create genuinely non-trivial equilibrium behaviors.

The network consists of multiple node types: Raw Material Suppliers $(S_1, S_2, S_3)$, Component Manufacturers $(M_1, M_2, M_3)$, Assembly Plants $(A_1, A_2)$, Testing Facilities $(T_1, T_2)$, and Distribution Centers $(D_1, D_2)$. The network topology includes several important cycles, particularly $M_1 \to A_1 \to T_1 \to M_1$ and $M_2 \to A_2 \to T_2 \to M_2$ (quality feedback loops), $M_1 \to A_2 \to T_2 \to M_1$ and $M_3 \to A_1 \to T_1 \to M_3$ (cross-department feedback), and $T_1 \leftrightarrow T_2$ (testing facilities sharing information): see Figure \ref{fig:QUIVERS}[right].

For each node $v$ in this network, the payment lattice is defined as:
\[
L_v = [0, C_v] \times [0,1]^{\mathcal{A}}
\]
where $[0, C_v]$ represents material quantities bounded by capacity $C_v$, and $\mathcal{A} = \{a_1, a_2, a_3\}$ corresponds to quality attributes: $a_1$ for durability, $a_2$ for precision, and $a_3$ for safety compliance. These attributes combine using G\"odel operations:
\[
x \otimes y = \min(x, y) 
\quad : \quad 
x \rightarrow y = 
\begin{cases}
1 & \text{if } x \leq y \\
y & \text{otherwise}
\end{cases}
\]

Unlike the financial compliance example where \L{}ukasiewicz operations model accumulating risk, the G\"odel operations model the "weakest link principle" that governs manufacturing quality. The key insight is that in manufacturing contexts, a product's overall quality is fundamentally limited by its least satisfactory attribute.

What makes this example non-trivial is the interplay between adaptive quality thresholds, quality-dependent production functions, testing nodes with memory effects, and cross-cutting material flows. We formalize these interactions through specialized operators at each node type.

First, each production node maintains a vector of minimum quality thresholds $\theta_v = (\theta_v^1, \theta_v^2, \theta_v^3)$ that evolve based on testing feedback:
\[
\theta_v^{(t+1)} = \min(\theta_v^{(t)}, q^{\text{feedback}} + \delta)
\]
where $\delta$ is a small tolerance vector. This creates path-dependent dynamics where quality standards adapt to network-wide performance.

Second, the pay-in aggregator at manufacturing nodes incorporates both quality combination and threshold-dependent production scaling:
\[
\inagg_v\left(\left\{(q_e, q_e')\right\}_{e \in t^{-1}(v)}\right) = \left(g_v\left(\sum_{e \in t^{-1}(v)} q_e, \bigotimes_{e \in t^{-1}(v)} q_e'\right), \bigotimes_{e \in t^{-1}(v)} q_e'\right)
\]
where the production function $g_v$ implements quality-dependent capacity:
\[
g_v(q, q') = q \cdot \min_{i}\left\{\max\left(0, \frac{q'_i - \theta_v^i}{1 - \theta_v^i}\right)^{\gamma}\right\}
\]
with parameter $\gamma > 1$ controlling how severely production drops when quality barely exceeds thresholds.

Third, testing facilities implement memory effects in their aggregators:
\[
\inagg_{T_j}\left((q, q')\right) = \left(0, \min(q', \omega \otimes q^{\text{previous}})\right)
\]
where $\omega \in (0,1)$ is a memory weight and $q^{\text{previous}}$ represents prior quality measurements. This creates temporal dependencies that complicate equilibrium finding.

The distributor functions further enhance network complexity. For a manufacturer $M_j$ connected to multiple assembly plants and receiving feedback from testing facilities:
\[
\dist_{M_j}(q, q') = \left(\{(q \cdot \pi_e, q')\}_{e \in s^{-1}(M_j) \cap \{A_1, A_2\}}, \{(0, q')\}_{e \in s^{-1}(M_j) \cap \{T_1, T_2\}} \right)
\]
where $\pi_e$ represents distribution proportions. Testing facilities distribute pure information:
\[
\dist_{T_j}(0, q') = \left\{(0, q')\right\}_{e \in s^{-1}(T_j)}
\]

This intricate system creates clearing sections with several distinctive properties:

First, multiple equilibria may exist depending on initial quality thresholds. Two networks with identical resources and capacity constraints but different starting quality standards can converge to different clearing sections, demonstrating genuine path dependency.

Second, quality bottlenecks propagate non-linearly through the network. When a quality attribute at one node approaches its threshold, the G\"odel minimum operation causes a sharp production decrease that reverberates through the network, creating ripple effects far from the original quality issue.

Third, feedback delays created by testing memory effects mean that clearing sections represent temporally stable configurations where all quality signals have fully propagated and been incorporated into production decisions.

Consider a concrete example within this network: Manufacturer $M_1$ receives raw materials from Supplier $S_1$ with quality profile $(0.9, 0.85, 0.95)$ and feedback from Testing Facility $T_1$ with quality profile $(0.85, 0.92, 0.88)$. The G\"odel aggregation produces:
\[
(0.9, 0.85, 0.95) \otimes (0.85, 0.92, 0.88) = (0.85, 0.85, 0.88)
\]

If $M_1$ has thresholds $\theta_{M_1} = (0.8, 0.8, 0.85)$, the production scaling factor would be:
\[
\min\left\{\left(\frac{0.85-0.8}{0.2}\right)^{\gamma}, \left(\frac{0.85-0.8}{0.2}\right)^{\gamma}, \left(\frac{0.88-0.85}{0.15}\right)^{\gamma}\right\}
\]

With $\gamma = 2$, this equals approximately $0.04$, severely restricting output despite all quality measures exceeding thresholds. If $M_1$ supplies both Assembly Plants $A_1$ and $A_2$, this bottleneck propagates through the network, potentially causing complex cascade effects.

Computing a clearing section for this network requires determining:
\begin{itemize}
\item Stable material quantities at all nodes
\item Stabilized quality profiles throughout the network
\item Converged quality thresholds at all production nodes
\item Steady-state memory effects at testing nodes
\end{itemize}

The feedback loops, memory effects, and quality-quantity interactions makes it difficult to guess what the clearing sections will look like through simple path minimization. Yet, this example is still very simplistic compared to what is possible (different residuated lattices at every node, tracking different attributes).


Our framework's ability to accommodate different residuated logics within the same mathematical structure allows precise modeling of diverse application domains with appropriate semantic interpretations. While a single residuated logic might be insufficient for modeling complex real-world phenomena, our general approach to lattice liability networks provides the flexibility needed to select the right logic for each application.

\section{Access Control Networks}
\label{sec:permission}

Permission and authorization systems are ubiquitous in digital infrastructure, typically modeled using simple directed graphs where edges represent delegation relationships. Standard approaches in security engineering often employ straightforward graph-traversal algorithms that propagate permissions through iterative set-union operations until reaching a fixed point. While these techniques are efficient and sufficient for many practical applications, they largely operate without the mathematical guarantees or structural insights that a more formal framework could provide. Our lattice liability approach offers precisely such a framework, providing rigorous existence guarantees and structural insights while accommodating more complex permission scenarios that challenge traditional methods.

We begin by considering a permission domain where the lattice-theoretic perspective adds significant value: systems where permissions have rich internal structure, multiple delegation constraints, cross-organizational boundaries, and cyclical delegation patterns. Let $\permset = \{\perm_1, \perm_2, \ldots, \perm_m\}$ be a finite set of atomic permissions. For each vertex (entity) $v \in V$ in our network, the payment lattice is the power set $L_v = \pow{\permset} = 2^\permset$ ordered by subset inclusion, reflecting the natural hierarchy of permission assignments. Structurally, $L_v$ is a boolean lattice, with set union ($\cup$) as join, set intersection ($\cap$) as meet, $\emptyset$ as bottom element, and $\permset$ as top element. This lattice structure provides a natural framework for reasoning about permissions and their relationships -- a feature particularly valuable when permissions have complex interdependencies.

The quiver $Q = (V, E, s, t)$ in this context represents entities that can delegate permissions. Unlike traditional permission graphs where edges simply indicate ``can delegate to,'' our framework allows for two critical refinements. First, each edge $e \in E$ carries both a \emph{minimum required delegation} $\liab_e^{\min} \subseteq \permset$ and a \emph{maximum allowed delegation} $\liab_e \subseteq \permset$, where $\liab_e^{\min} \subseteq \liab_e$. The set $\liab_e^{\min}$ forms a downset in the permission lattice: if a permission $\perm$ must be delegated, any permission implied by $\perm$ must also be delegated. This dual constraint system supports sophisticated delegation policies impractical in simpler models. Second, our approach handles typed permissions with delegation depth limits through conversion functions, capturing complex security policies like ``permissions degraded after two delegation hops'' or ``higher privileges are non-transferable.''

For each vertex $v \in V$, the pay-in aggregator $\inagg_v: \prod_{e \in t^{-1}(v)} L_{s(e)} \to L_v$ computes the join of incoming permissions in the boolean lattice:

\begin{equation}
\inagg_v\left(\left(\epymt_e\right)_{e \in t^{-1}(v)}\right) 
= \bigcup_{e \in t^{-1}(v)} \conv_e(\epymt_e)
\end{equation}

where $\conv_e: L_{s(e)} \to L_v$ converts permissions as they cross organizational boundaries. This conversion function provides expressiveness beyond simple joins. For example, it can downgrade or modify transferrability. To handle permissions that cannot be delegated further, we augment our network with self-loops. For each vertex $v \in V$, we add a non-delegatable permissions loop $e_v^{\text{non-del}}: v \to v$ with nominal liability $\liab_{e_v^{\text{non-del}}} = \permset$. Similarly, exogenous resource loops $e_v^{\text{in}}: v \to v$ with $\liab_{e_v^{\text{in}}} = \iota_v \subseteq \permset$ capture inherent permissions that entities possess independently of any delegation.

The distributor $\dist_v: L_v \to L_v^{|s^{-1}(v)|}$ allocates permissions along outgoing edges:

\begin{equation}
[\dist_v(x_v)]_e = 
\begin{cases}
x_v \cap \liab_e & \text{if } e \in s^{-1}(v) \cap E \\
x_v \setminus \bigcup_{e' \in s^{-1}(v) \cap E} \liab_{e'} & \text{if } e = e_v^{\text{non-del}}
\end{cases}
\end{equation}

This ensures that entity $v$ delegates only permissions it actually possesses ($x_v$) and respects maximum allowed delegations ($\liab_e$). The corresponding pay-out aggregator $\outagg_v$ combines these distributed permissions using the join operation in the boolean lattice.

A clearing section in this context represents a network-wide permission assignment where each entity's permissions are consistent with both what it receives from others and what it delegates. Unlike financial networks where clearing sections represent payment flows, permission network clearing sections represent stable distributions of privileges. The collection of valid clearing sections forms an upset in the product lattice structure: if $\mathbf{x} = (x_v)_{v \in V}$ is a valid clearing section and $\mathbf{y} \geq \mathbf{x}$ in the product order, then $\mathbf{y}$ is also valid, as it satisfies all minimum delegation requirements:

\begin{equation}
[\dist_v(x_v)]_e \supseteq \liab_e^{\min} \text{ for all } e \in E
\end{equation}

This upset structure is precisely what allows us to identify the \emph{least-privilege clearing section} -- the meet of all valid clearing sections, representing the minimal permission assignment that satisfies all security requirements. This directly implements the principle of least privilege, a cornerstone of secure system design. Theorem~\ref{thm:existence} guarantees this least clearing section exists, and the distributed computation approach in Section~\ref{sec:computational} provides a practical method for computing it, starting from the minimal state that could potentially satisfy delegation requirements.

The lattice-theoretic perspective provides several unique insights into permission networks that traditional approaches may obscure. First, it reveals how permission cycles can amplify privilege requirements: if entities $v_1 \to v_2 \to \cdots \to v_n \to v_1$ form a cycle where each must delegate certain permissions to the next, all entities in the cycle may need to possess permissions required by any member of the cycle. This phenomenon, which we might call ``privilege amplification through cyclic delegation,'' represents a potential security risk that becomes immediately apparent in our framework.

Second, our approach illuminates the relationship between network topology and minimal permission requirements. The least clearing section quantifies precisely how network structure affects minimal privilege assignments, revealing which entities serve as permission amplifiers or bottlenecks. This supports principled security architecture decisions rather than ad-hoc permission assignments.

Finally, the lattice structure of clearing sections provides natural semantics for common security operations. For instance, privilege escalation assessment becomes a comparison between minimal required permissions and inherent permissions, while resilience testing can be formalized as analyzing how clearing sections change when removing entities or delegation paths.

While simpler approaches suffice for basic permission management scenarios, our lattice liability framework offers substantial advantages for complex, cross-organizational, or security-critical systems. Rather than proposing this as a replacement for existing authorization systems, we present it as a theoretical foundation that unifies permission management with the broader theme of network equilibria, demonstrating how seemingly disparate domains share fundamental mathematical structures. This unification perspective allows security architects to leverage insights from other domains (such as financial network stability) when designing robust permission systems.

\section{Generalizations to Chain-Complete Lattices}
\label{sec:chaincomplete}

Up to this point, we have primarily assumed that each vertex lattice $L_v$ in our network is a \emph{complete} lattice, where Tarski's Fixed Point Theorem directly guarantees the existence of clearing sections. However, many real-world applications naturally give rise to lattices that are only \emph{chain-complete}. In this section, we show how our framework extends to such settings, laying the groundwork for applications involving infinite-dimensional state spaces such as the function spaces we'll encounter in term structure models.

Complete lattices provide a convenient setting for fixed-point theorems, but they can be restrictive when modeling certain domains. Consider the set of continuous functions $f: [0,1] \to \mathbb{R}$ with pointwise ordering. While suprema of chains (totally ordered subsets) exist and remain continuous, the pointwise supremum of an arbitrary collection of continuous functions may not be continuous. Similarly, in systems with uncertainty, state spaces often involve probability distributions where arbitrary suprema might not maintain the constraints of probability measures. When resources flow over time with temporal constraints (as in scheduled payments or production timelines), the appropriate state spaces are function spaces that are rarely complete lattices but often satisfy chain-completeness.

Such examples illustrate a common pattern: as we move beyond simple scalar-valued resources to more structured state spaces, we often lose completeness while retaining chain-completeness. This property turns out to be sufficient for our framework, provided we impose appropriate continuity conditions.

A partially ordered set $(P, \le)$ is called \emph{chain-complete} if for every totally ordered subset (chain) $C \subseteq P$, both $\sup C$ and $\inf C$ exist in $P$. Appendix~\ref{sec:chains} provides a comprehensive treatment of chain-complete partially ordered sets and their properties.

To extend our framework to chain-complete lattices, we require a few modifications to Definition \ref{def:LLN}:
\begin{enumerate}
\item Assume each vertex lattice $L_v$ is \emph{chain-complete} rather than complete.
\item Assume all aggregator and distributor maps preserve suprema of chains (Scott continuity).
\item Modify Equation \ref{eq:Bounded-Liability} to encode bounded liability without invoking a specific $\top$-element in each payment lattice.
\end{enumerate}

With these modifications in place and Definition \ref{def:clearing-section} unchanged, Theorem \ref{thm:existence} can be generalized as follows.

\begin{theorem}[Clearing Sections for Chain-Complete Systems]
\label{thm:chaincomplete-existence}
Let $\LLN_Q$ be a lattice liability network for which all lattices $L_v$ are chain-complete and all aggregator and distributor maps are chain-sup-preserving.  Then:
\begin{enumerate}
\item There exists at least one clearing section.
\item The set of all clearing sections is chain-complete under the product order.
\end{enumerate}
\end{theorem}

{\em Proof:}
By assumption, each $L_v$ is chain-complete, and there are finitely many vertices $v\in V$.  Hence the product $C^0 = \prod_{v\in V}\, L_v$ is chain-complete.  The global update map $\Phi$ is monotone and preserves suprema of chains by composition.  By
the Kleene-Tarski Theorem (see Theorem~\ref{thm:chainTarski} in Appendix \ref{sec:chaincomplete}), $\Phi$ has at least one fixed point.  Thus a clearing section exists.

To see that $\Fix(\Phi)$ forms a chain-complete set, let $\{\mathbf{x}_\alpha\}$ be a chain of clearing sections in $C^0$. Its coordinatewise supremum $\mathbf{x} = \sup_\alpha\,\mathbf{x}_\alpha$
lies in $C^0$.  By chain-sup-preservation of $\Phi$,
\[
\Phi(\mathbf{x})
= \Phi\!\Bigl(\sup_\alpha\,\mathbf{x}_\alpha\Bigr)
= \sup_\alpha\,\Phi\bigl(\mathbf{x}_\alpha\bigr)
= \sup_\alpha\,\mathbf{x}_\alpha
= \mathbf{x}.
\]
So $\mathbf{x}$ is again a clearing section.  Thus the set of all clearing sections is closed under suprema of chains, and a dual argument shows it is also closed under their infima.  Therefore the
collection of all clearing sections is a chain-complete sub-poset of $\prod_{v\in V} L_v$.
\qed

The extension to chain-complete systems significantly broadens the applicability of our framework. We can now model payment flows, production rates, or resource allocations that vary continuously over time or space -- essential for the term structure models to be examined in Section~\ref{sec:termstructure}, where payment obligations are represented as functions over a time interval. More generally, we can handle infinite-dimensional state spaces that capture continuous variations in multiple dimensions, such as spatiotemporal resource distributions or parameter-dependent obligations.

Chain-completeness also allows us to model systems with potentially unbounded growth or decay, where complete lattices might impose artificial bounds. From a computational perspective, the chain-sup-preservation condition aligns naturally with iterative methods, as it ensures that approximation sequences converge to proper fixed points.

\begin{remark}[Distributed Computation in Chain-Complete Settings]
The distributed algorithm presented in Section~\ref{sec:computational} extends naturally to chain-complete systems. When using the modified iteration framework of Theorem~\ref{thm:chaincomplete-existence}, nodes must employ Scott-continuous aggregators and distributors. The iterative process still converges, though termination detection becomes more nuanced. For infinite lattices, nodes may need to detect when consecutive states are sufficiently close according to an appropriate topology rather than exactly equal.
\end{remark}

The only additional hypothesis we have required beyond monotonicity -- that our operators preserve suprema of chains -- is quite natural in practice. Most aggregators and distributors in applications are built from operations (sums, integrals, pointwise operations) that commute with taking suprema of chains. This generalization maintains the conceptual elegance of our framework while significantly extending its reach to the richer state spaces needed for sophisticated applications like term structure models, continuous-time dynamic systems, and stochastic networks.

\section{Term Structure Models}
\label{sec:termstructure}

Financial instruments are inherently temporal: bonds, loans, and other debt instruments promise payments at future dates. The \emph{yield curve} (or term structure of interest rates) captures this temporal dimension by mapping maturity times to interest rates. We now demonstrate how our framework elegantly handles networks where obligations depend on term structures, and where completeness fails but chain-completeness suffices as described in Section~\ref{sec:chaincomplete}.

Consider a network of financial institutions connected through various maturity-dependent obligations. Each institution's state is characterized not by a single payment amount but by a complete yield curve. This naturally gives rise to lattices that are chain-complete but not complete.

\begin{definition}[Term Structure Lattice]
For each vertex $v \in V$, define the payment lattice $L_v$ as:
\[
L_v = \left\{f: [0,T] \to \mathbb{R}_{\geq 0} \;\middle|\; f \text{ is right-continuous and bounded}
\right\}
\]
equipped with pointwise ordering: $f \leq g$ if and only if $f(t) \leq g(t)$ for all $t \in [0,T]$.
\end{definition}

This lattice $L_v$ represents all possible payment schedules over the time horizon $[0,T]$, where each function $f \in L_v$ specifies the payment rate at time $t$. The integral constraint ensures finite total payments. Unlike the product lattices of intervals seen in previous examples, $L_v$ is only chain-complete, not complete: while the pointwise supremum of any chain of functions in $L_v$ remains in $L_v$ (by the monotone convergence theorem), arbitrary collections of functions may have suprema that violate boundedness or integrability conditions.

The nominal liabilities between institutions are now function-valued:

\begin{definition}[Term-Structured Liabilities]
For each edge $e \in E$, the nominal liability is a function $\liab_e \in L_{s(e)}$ representing the payment schedule from $s(e)$ to $t(e)$.
\end{definition}

The pay-in aggregator at vertex $v$ must combine incoming payment schedules. Given the temporal nature of payments, this aggregation must account for the time value of money, maturity transformation, and discount factors:

\begin{definition}[Term-Structured Pay-in Aggregator]
For each vertex $v \in V$, the pay-in aggregator $\inagg_v: \prod_{e \in t^{-1}(v)} L_{s(e)} \to L_v$ is defined by:
\[
[\inagg_v(\{f_e\}_{e \in t^{-1}(v)})](t) = \sum_{e \in t^{-1}(v)} \conv_e(f_e)(t)
\]
where $\conv_e: L_{s(e)} \to L_v$ is a term structure conversion operator:
\[
[\conv_e(f)](t) = \int_0^T K_e(t,s)f(s)\,ds
\]
with kernel $K_e(t,s) \geq 0$ representing both conversion rates, temporal redistribution, and discount factors.
\end{definition}

The conversion operator $\conv_e$ transforms payments from the time domain of the source institution to that of the target institution. The kernel $K_e(t,s)$ can incorporate exchange rates, credit risk premiums, and term premia. For simple conversions, $K_e(t,s) = \rho_e \cdot \updelta(t-s)$ where $\rho_e$ is a conversion rate and $\updelta$ is the Dirac delta function. For maturity transformation (borrowing short to lend long), $K_e$ would have significant off-diagonal elements, representing how short-term funding contributes to long-term obligations.

The distributor-aggregator pair handles the allocation of payment capacity across time and counterparties:

\begin{definition}[Term-Structured Distributor-Aggregator]
For each vertex $v \in V$, the distributor $\dist_v: L_v \to L_v^{|s^{-1}(v)|}$ is defined by:
\[
[\dist_v(f)]_e(t) = \min\left\{\liab_e(t), \pi_e(t) \cdot f(t)\right\}
\]
where $\pi_e(t)$ is a time-varying proportional factor:
\[
\pi_e(t) = \frac{\liab_e(t)}{\sum_{e' \in s^{-1}(v)} \liab_{e'}(t)}
\]

The corresponding pay-out aggregator $\outagg_v: L_v^{|s^{-1}(v)|} \to L_v$ is defined by:
\[
[\outagg_v(\{g_e\}_{e \in s^{-1}(v)})](t) = \sum_{e \in s^{-1}(v)} g_e(t)
\]
\end{definition}

This distributor implements a time-specific proportional payment rule: at each maturity $t$, obligations are paid proportionally to their due amounts, up to the nominal liability. The aggregator simply sums all distributed payments across counterparties at each time point.

\begin{lemma}
The term structure operators satisfy the required properties for Theorem~\ref{thm:chaincomplete-existence}:
\begin{enumerate}
\item The pay-in aggregator $\inagg_v$ and distributor $\dist_v$ are monotone
\item The distributor $\dist_v$ respects liability bounds: $[\dist_v(f)]_e(t) \leq \liab_e(t)$ for all $t \in [0,T]$
\item The factorization identity $\outagg_v \circ \dist_v = \id_{L_v}$ holds when capacity is sufficient to meet all liabilities
\item Both $\inagg_v$ and $\dist_v$ preserve suprema of chains (Scott-continuity)
\end{enumerate}
\end{lemma}

{\em Proof:} Monotonicity follows from the properties of integration with positive kernels and the monotonicity of the minimum function. Liability bounds are respected by construction of the $\min$ operation in the distributor. The factorization identity holds because $\sum_{e \in s^{-1}(v)} \pi_e(t) = 1$ when defined. 

Chain-supremum preservation follows from the monotone convergence theorem: if $\{f_n\}$ is an increasing chain of functions in $L_v$, then 
\[
    \int K_e(t,s) \sup_n f_n(s) ds = \sup_n \int K_e(t,s) f_n(s) ds ,
\] 
when $K_e \geq 0$. Similarly, for the distributor, 
\[
    \min\{\liab_e(t), \pi_e(t) \cdot \sup_n f_n(t)\} = \sup_n \min\{\liab_e(t), \pi_e(t) \cdot f_n(t)\} .
\]
Thus, both operators preserve suprema of chains.
\qed

A clearing section in this context represents a consistent set of payment schedules across all institutions that satisfies obligations at each point in time. For each institution $v$, the clearing function $x_v \in L_v$ specifies its payment rate at each time point. This models how financial institutions manage their temporal cash flows to meet ongoing obligations while respecting their own funding constraints.

By Theorem~\ref{thm:chaincomplete-existence}, a clearing section exists in this term structure network because our lattices are chain-complete and our operators preserve suprema of chains. Such a clearing section represents a consistent set of payment schedules that respects temporal constraints, conversion effects, and payment priorities over the entire time horizon.

\begin{remark}[Computation via Discretization]
Computing clearing sections in term structure models requires discretizing the time domain $[0,T]$ due to their infinite-dimensional nature. Practical implementations represent functions using finite basis expansions (splines or wavelets) with updates performed in coefficient space. Convergence detection employs functional distance measures such as $L^2$ or $L^\infty$ norms. Notably, longer-term obligations typically converge more slowly than short-term ones, creating a ``term structure of convergence rates'' that mirrors real-world market behavior where near-term rates equilibrate faster due to greater certainty in short-term obligations and funding conditions.
\end{remark}

\begin{remark}[Comparison with Optimal Control Approach]
Our term structure formulation contrasts with Barratt and Boyd's \cite{barratt2020multi} multi-period liability clearing model. Our continuous formulation offers advantages for modeling yield curves, maturity transformations, and temporal discount factors, while their approach handles discrete scheduling constraints more naturally and is typically more computationally tractable. An interesting research direction would be combining these approaches, merging convex control and lattice methods.
\end{remark}

Our framework naturally accommodates richer structures where maturity transformations (borrowing short to lend long) create complex temporal dependencies. For instance, the conversion operator $\conv_e$ could incorporate liquidity premia or maturity transformation costs, modeled as off-diagonal elements in the kernel $K_e(t,s)$. This allows analyzing how disruptions in short-term funding markets can cascade into longer-term obligations and vice versa, a key mechanism in systemic financial crises.

\section{Multivalued Extensions \& Uncertainty}
\label{sec:multival}

Real-world networks routinely operate under conditions of uncertainty, imperfect information, or inherent variability. Market prices fluctuate, demand varies stochastically, and agents may have multiple viable strategies rather than unique optimal responses. The lattice liability framework developed thus far assumes deterministic, single-valued mappings between lattice elements. We now extend this framework to accommodate multivalued relationships, allowing us to model uncertainty, noise, and strategic ambiguity while maintaining the fundamental lattice-theoretic structure.

For the sake of computational tractability, we will restrict to the case of finite lattices (automatically complete).  Given finite lattices $L$ and $M$, we define monotonicity properties for set-valued mappings of $L$ to the powerset lattice $\pow{M}$:

\begin{definition}[Monotone Correspondences]
A correspondence $F: L \to \pow{M}$ is:
\begin{itemize}
\item \emph{Upper monotone} if for all $x_1 \leq x_2$ in $L$ and all $y_1 \in F(x_1)$, there exists $y_2 \in F(x_2)$ with $y_1 \leq y_2$.
\item \emph{Lower monotone} if for all $x_1 \leq x_2$ in $L$ and all $y_2 \in F(x_2)$, there exists $y_1 \in F(x_1)$ with $y_1 \leq y_2$.
\item \emph{Monotone} if it is both upper and lower monotone.
\end{itemize}
\end{definition}

Upper monotonicity ensures that higher inputs can produce at least equally high outputs, while lower monotonicity ensures that lower outputs are possible from lower inputs. Together, these properties preserve the essential order structure while allowing for one-to-many relationships.

\begin{definition}[Multi-Valued Lattice Liability Network]
A \emph{multi-valued lattice liability network} consists of:
\begin{enumerate}
\item For each vertex $v \in V$, a finite lattice $L_v$;

\item For each edge $e \in E$, a nominal liability $\liab_e \in L_{s(e)}$ (as in the single-valued case), bounding the maximum payment along edge $e$;

\item For each vertex $v \in V$, a monotone pay-in correspondence $\inagg_v: \prod_{e \in t^{-1}(v)} L_{s(e)} \to \pow{L_v}$;

\item For each vertex $v \in V$, a distribution-aggregation pair $(\dist_v, \outagg_v)$ where:
   \begin{itemize}
   \item $\dist_v: L_v \to \pow{L_v^{|s^{-1}(v)|}}$ is a monotone distributor correspondence;
   \item $\outagg_v: L_v^{|s^{-1}(v)|} \to \pow{L_v}$ is a monotone aggregator correspondence;
   \item For all $x_v \in L_v$, all $\mathbf{y} \in \dist_v(x_v)$, and all $e \in s^{-1}(v)$, we have $y_e \leq \liab_e$ (respecting liability bounds)
   \item For all $x_v \in L_v$ and all $\mathbf{y} \in \dist_v(x_v)$, there exists $z \in \outagg_v(\mathbf{y})$ with $z = x_v$ (resource conservation)
   \end{itemize}
\end{enumerate}
\end{definition}

The key difference from the single-valued case is that the aggregators and distributors now return sets of possible values rather than single values. The nominal liabilities, however, remain fixed single elements as in the original framework. This preserves the conceptual clarity of fixed obligations while introducing flexibility in how resources are aggregated and distributed.

The liability bound condition ensures that each possible distribution respects the maximum payment constraints along each edge. The resource conservation condition is a generalized form of the identity factorization requirement from the single-valued case: it ensures that for any resource state, at least one possible distribution and re-aggregation path preserves the original resources.

Unlike the single-valued case where each node has a unique state, a multi-valued clearing section assigns a set of possible states to each vertex. These sets represent ranges of plausible outcomes under uncertainty.

\begin{definition}[Multi-Valued Clearing Section]
A \emph{multi-valued clearing section} is an assignment $\mathbf{X} = (X_v)_{v \in V}$ where each $X_v \subseteq L_v$ is nonempty, such that:
\begin{enumerate}
\item For each vertex $v \in V$ and each $x_v \in X_v$, there exist selections $x_{s(e)} \in X_{s(e)}$ for all $e \in t^{-1}(v)$ such that
$x_v \in \inagg_v\left(\{x_{s(e)}\}_{e \in t^{-1}(v)}\right)$

\item For each vertex $v \in V$ and each $x_v \in X_v$, there exists $\mathbf{y} \in \dist_v(x_v)$ such that $y_e \in X_{t(e)}$ for all $e \in s^{-1}(v)$
\end{enumerate}
\end{definition}

This definition ensures mutual consistency between the sets of states. For any state in a node's set, there must exist compatible states in neighboring nodes that could produce it (condition 1) and that it could help produce (condition 2). This two-way consistency captures the essence of network equilibrium in an uncertain setting.

To establish existence of multi-valued clearing sections, we define an operator $\Phi$ on the complete lattice $C^0 = \prod_{v \in V} \pow{L_v}$ (the product of power sets of vertex lattices). For a given $\mathbf{X} = (X_v)_{v \in V} \in C^0$, we define:

\[
[\Phi(\mathbf{X})]_v = \left\{y \in L_v : \begin{array}{l}
y \in \inagg_v\left(\{x_{s(e)}\}_{e \in t^{-1}(v)}\right) \text{ for some } x_{s(e)} \in X_{s(e)}, \\
\text{and there exists } \mathbf{z} \in \dist_v(y) \text{ with } z_e \in X_{t(e)} \text{ for all } e \in s^{-1}(v)
\end{array} \right\}
\]

We can partially order $C^0$ by declaring $\mathbf{X} \leq \mathbf{Y}$ if for each vertex $v$ and each $x \in X_v$, there exists $y \in Y_v$ with $x \leq y$, and for each $y \in Y_v$, there exists $x \in X_v$ with $x \leq y$. This ordering makes $C^0$ a complete lattice.

\begin{theorem}[Existence of Multi-Valued Clearing Sections]
Let $\mathcal{M}_Q$ be a multi-valued lattice liability network as defined above. Then:
\begin{enumerate}
\item There exists at least one multi-valued clearing section
\item The set of all multi-valued clearing sections forms a complete lattice under the induced order defined above
\end{enumerate}
\end{theorem}

{\em Proof:} Under the given conditions, the operator $\Phi$ is monotone on the complete lattice $C^0$. Moreover, for each $\mathbf{X} \in C^0$, the set $\Phi(\mathbf{X})$ has a $\leq$-maximal element and $\leq$-minimal element. By Zhou's fixed point theorem for correspondences \cite{Zhou1994,Yu2024}, $\Phi$ has a fixed point, and the set of all fixed points forms a complete lattice.
\qed

This multivalued framework offers several advantages for modeling real-world systems:

\begin{enumerate}
\item \emph{Robustness analysis}: By capturing ranges of possible states rather than point estimates, we can assess how robust network equilibria are to various perturbations.

\item \emph{Uncertainty propagation}: The framework naturally tracks how uncertainty in one part of the network affects other parts, revealing which nodes act as uncertainty amplifiers or dampeners.

\item \emph{Strategic indeterminacy}: When agents have multiple best responses or operate under bounded rationality, the multivalued approach captures the resulting range of potential outcomes.

\item \emph{Scenario bounding}: Even when exact solutions cannot be determined, the framework identifies upper and lower bounds on possible system states, supporting risk assessment and contingency planning.
\end{enumerate}

\begin{remark}
The distributed computation approach of Section~\ref{sec:computational} generalizes to multivalued settings, albeit with important modifications. Instead of computing a single state update, each node must now compute a set of possible states in each iteration. A natural implementation is to track the \emph{boundaries} of these state sets -- their minimal and maximal elements -- rather than enumerating all possible states. For finite lattices, this boundary-tracking approach has the same asymptotic complexity as the single-valued case. The termination protocol becomes more complex, as nodes must detect when their state-set boundaries have stabilized rather than single states. Uncertainty propagation can be visualized by tracking how the ``width'' of state sets (distance between minimal and maximal elements) evolves through the network, identifying nodes that amplify or dampen uncertainty.
\end{remark}

This approach to handling uncertainty preserves the mathematical elegance of the lattice liability framework while substantially increasing its practical applicability to real-world systems characterized by noise, ambiguity, and strategic complexity.

\section{Complex Negotiation Networks}
\label{sec:negotiation}

Negotiation processes represent a compelling application of our multivalued framework, as they naturally involve strategic flexibility, competing objectives, and multiple potential outcomes. Unlike financial transactions where value maximization provides a clear ordering principle, negotiations often involve inherent tradeoffs that cannot be reduced to simple numerical comparisons. We demonstrate how finite partial orders can effectively model this complexity while maintaining mathematical tractability.

Consider a network of parties engaged in interconnected negotiations. The fundamental challenge in modeling negotiations is capturing the intrinsic tradeoffs that make certain positions incomparable. For instance, in a price-quantity negotiation, a seller might view $(€15, 150\text{ units})$ and $(€12, 200\text{ units})$ as incomparable alternatives -- the former offers higher per-unit revenue while the latter promises greater volume. Neither position dominates the other in all dimensions, creating a partially ordered preference structure rather than a simple linear ordering.

To capture this richness, we model each party's negotiation positions using a finite partial order rather than a product of linear orders. Formally, for a quiver $Q = (V, E, s, t)$ where vertices represent negotiating parties, we construct for each party $v \in V$ a finite partial order $(P_v, \leq_v)$ whose elements represent possible negotiation positions. The partial order relation captures clear preference: $p \leq_v q$ means position $q$ is unambiguously preferred to position $p$ by party $v$. Crucially, many positions remain incomparable, reflecting genuine tradeoffs rather than indifference.

For each partial order $P_v$, we can construct its corresponding lattice of downsets $L_v = \mathcal{O}(P_v)$, where each element of $L_v$ represents a downward-closed subset of positions. Intuitively, an element $X \in L_v$ represents a claim of the form ``Party $v$ can achieve at least one of the positions in set $X$.'' The lattice order corresponds to strength of negotiating position: $X \leq Y$ means $Y$ represents a stronger position than $X$ because it guarantees at least the same outcomes and potentially better ones. This construction transforms our partial orders into finite distributive lattices -- precisely the mathematical structure our framework requires.

Each party in a negotiation network also has external factors influencing its position -- alternatives outside the negotiation, resource constraints, or legal requirements. Following \S\ref{sec:exo}, we augment each vertex with exogenous resource loops. For each vertex $v \in V$, we add a self-loop $e_v^{\text{in}}$ with $\liab_{e_v^{\text{in}}} = \iota_v \in L_v$, representing the party's BATNA (Best Alternative To Negotiated Agreement) and other external constraints. These exogenous resources establish a baseline position that the party can maintain regardless of negotiation outcomes. Importantly, the self-loop feeds directly into the pay-in aggregator $\inagg_v$, allowing the BATNA to be naturally incorporated alongside other incoming offers when determining the party's position set.

For each edge $e \in E$ from party $v$ to party $w$, we define a maximal concession bound $\liab_e \in L_{s(e)}$ that represents the most $v$ would ever be willing to concede to $w$. For example, if party $v$ can offer at most position set $Z \in L_v$ to party $w$ (perhaps due to resource limitations or strategic considerations), then $\liab_e = Z$ defines this bound. Unlike financial liabilities which represent obligations that must be met, these negotiation bounds represent upper limits on concessions that parties would prefer not to reach and certainly never exceed. This reinterpretation of the nominal liability as a maximal concession aligns naturally with negotiation dynamics where parties set boundaries on their flexibility.

The operational components of our multi-valued lattice liability network take on specific meanings in the negotiation context:

\begin{enumerate}
\item 
The pay-in aggregator correspondence $\inagg_v: \prod_{e \in t^{-1}(v)} L_{s(e)} \to \pow{L_v}$ captures how party $v$ incorporates incoming offers to update its internal position. Given offers $Y_1, Y_2, \ldots$ from different counterparties, $\inagg_v(Y_1, Y_2, \ldots)$ is the set of all internally consistent positions that party $v$ might adopt in response to these offers. This might involve strengthening its position if receiving favorable offers from multiple sources (e.g., $v$ might raise its asking price if multiple buyers show interest), or making internal concessions if all incoming offers are firm. 

\item 
The distributor correspondence $\dist_v: L_v \to \pow{L_v^{|s^{-1}(v)|}}$ maps a party's internal position to possible sets of outgoing offers. If party $v$ holds position set $X_v$, then $\dist_v(X_v)$ represents all possible combinations of counteroffers that $v$ might extend to its negotiating partners, constrained by $v$'s current position and the nominal liability bounds. This captures strategic decisions about making different concessions to different parties. For example, a supplier negotiating with multiple buyers might offer better terms to a strategic long-term customer while taking a firmer stance with occasional purchasers.

\item 
The pay-out aggregator correspondence $\outagg_v: L_v^{|s^{-1}(v)|} \to \pow{L_v}$ ensures that outgoing offers remain consistent with the internal position. The resource conservation property requires that for each $X_v \in L_v$ and counteroffer vector $\mathbf{Y} \in \dist_v(X_v)$, there exists $Z \in \outagg_v(\mathbf{Y})$ with $X_v \subseteq Z$. This means that if party $v$ makes offers $Y_1, Y_2, \ldots$ to its counterparties, these offers must be reconcilable with $v$'s internal position -- the party cannot promise more than it can deliver across all its negotiations.
\end{enumerate}

A key concept in negotiation theory is the \emph{zone of possible agreement} (ZOPA) -- positions acceptable to both parties. In our framework, for an edge $e$ from party $v$ to party $w$, the ZOPA consists of positions that lie in both $\dist_v(X_v)_e$ for some $X_v \in L_v$ and $\{Y \in L_{s(e)} : Y \in \inagg_w^{-1}(X_w) \text{ for some } X_w \in L_w\}$. Negotiations are viable only when ZOPAs are non-empty.

A multivalued clearing section $\mathbf{X} = (X_v)_{v \in V}$ in this context represents a network-wide ``equilibrium'' of negotiating positions where each party maintains a set of possible stances that are mutually compatible with all other parties' sets of possible stances. Unlike financial clearing where exact amounts must be determined, negotiation clearing involves ranges of positions that could potentially lead to agreement. This captures the fundamental uncertainty and strategic flexibility inherent in negotiations: parties don't commit to single positions but rather maintain sets of acceptable positions that are collectively viable. 

It is worth noting that while financial networks require resource conservation (with overflow loops handling excess), negotiations have fewer conservation requirements for bargaining power or strategic flexibility. Unused bargaining capacity simply remains as potential that could be exercised in future negotiation rounds. The absence of strict conservation principles is precisely what makes the multi-valued framework especially suitable for negotiation contexts.

Formally, a clearing section satisfies:
\begin{enumerate}
\item 
For each vertex $v \in V$ and each $x_v \in X_v$, there exist selections $x_{s(e)} \in X_{s(e)}$ for all $e \in t^{-1}(v)$ such that $x_v \in \inagg_v\left(\{x_{s(e)}\}_{e \in t^{-1}(v)}\right)$ - meaning each position in a party's position set is a possible response to some combination of positions from its counterparties.

\item 
For each vertex $v \in V$ and each $x_v \in X_v$, there exists $\mathbf{y} \in \dist_v(x_v)$ such that $y_e \in X_{t(e)}$ for all $e \in s^{-1}(v)$ - meaning each position in a party's position set can generate offers that are acceptable within the position sets of its counterparties.
\end{enumerate}

The multi-valued framework is particularly well-suited to negotiations precisely because negotiations inherently involve ranges of acceptable outcomes rather than single fixed positions. Unlike financial transactions where exact payment amounts are required, negotiations proceed through the exploration of possible agreement spaces, with parties maintaining strategic flexibility until final terms are fixed. By modeling negotiation equilibria as sets of mutually compatible positions rather than single positions, we capture the essential strategic flexibility that characterizes real negotiation processes.

The existence of clearing sections follows directly from our main theorem for multivalued lattice liability networks, since the position lattices $L_v$ are finite distributive lattices and all correspondence operations preserve the required monotonicity properties. Moreover, the set of all multivalued clearing sections forms a complete lattice under the appropriate ordering, with the greatest clearing section identifying the broadest range of mutually compatible positions and the least clearing section identifying the minimum necessary agreements.

This lattice structure of clearing sections reveals important insights about negotiation dynamics:
\begin{enumerate}
\item 
Larger sets in a clearing section indicate greater strategic flexibility. When a party's position set contains many alternatives, it has more maneuverability in responding to counterparties' moves. The size of position sets thus provides a formal measure of bargaining power.

\item 
The existence of multiple clearing sections reveals path dependence in negotiations. Different initial positions or negotiation sequences can lead to different equilibrium outcomes, even with identical preference structures. This models how anchoring effects and framing can influence final agreements.

\item 
The lattice structure shows how agreements in one part of the network constrain possibilities elsewhere. If parties $v$ and $w$ reach an agreement that restricts their positions, this propagates through the network, potentially limiting options for other parties. This captures how early agreements in multi-party negotiations can shape the overall solution space.
\end{enumerate}

This approach to negotiation networks demonstrates the flexibility of our multivalued framework in \S\ref{sec:multival}. By modeling the inherently multidimensional, strategic nature of negotiations through partial orders and their corresponding lattices of downsets, we capture the complex preference structures that make real negotiations challenging. The multivalued aspect reflects both strategic flexibility and inherent uncertainty, while the lattice structure provides the mathematical foundation for analyzing equilibrium outcomes and their properties.

\section{Conclusions}
\label{sec:conc}

This paper has introduced lattice liability networks as a unifying generalization of the Eisenberg-Noe framework for financial clearing. By recasting network clearing problems in the language of lattice-enriched quivers, we have developed a mathematical framework that preserves the fixed-point structure of the original model while substantially expanding its expressive power. Our main theorem establishes that under appropriate monotonicity conditions, clearing sections exist and form a complete lattice, providing both theoretical guarantees about equilibria and practical insights into system behavior.

The framework's power lies in its flexibility. We have demonstrated its applicability to diverse domains: multi-currency financial systems with non-linear conversion effects; decentralized finance systems with automated market makers; manufacturing networks with resource transformation; fuzzy payment systems capturing qualitative obligations; and complex multi-attribute negotiations. The generalization to chain-complete lattices further extends our framework to settings with infinite-dimensional state spaces, such as term structure models with continuous yield curves.

Several promising directions for future research emerge from this work. First, incorporating stochastic elements represents the most pressing extension. Real-world networks operate under significant uncertainty, with volatile asset values, random payment timing, and probabilistic default events. Developing a stochastic version of our framework would bridge the gap between our deterministic model and the inherent randomness of financial systems. Second, tailoring the framework to more realistic financial settings, including heterogeneous financial institutions with complex strategic behaviors, regulatory constraints, and market frictions, would enhance its practical applicability. Third, developing efficient computational methods for finding clearing sections in large-scale networks remains an important challenge, particularly for applications involving high-dimensional lattices or complex distributor functions.

The lattice liability network framework illustrates how order-theoretic structures can illuminate the fundamental properties of interconnected systems. By revealing common mathematical patterns behind seemingly different network phenomena, we provide both a theoretical foundation for understanding complex systems and a practical toolkit for analyzing their behavior across multiple domains.


\section*{Acknowledgments}

Several of the authors [RG, JG, ML] were supported by the Air Force Office of Scientific Research under award number FA9550-21-1-0334. 

The inspiration for this project came in a series of conversations between RG and Claude-3.5 in April 2024. Many of the application sections were devised with the assistance of Claude-3.5, Claude-3.7, and GPT-o1-pro. In addition, GPT-o3 [Deep Research] was extremely useful in generating context and assisting with the literature search. All figures are by RG. To an extent, this paper was an experiment in collaborative research using large language models.

\bibliographystyle{plain}
\bibliography{LLN}

\begin{thebibliography}{10}

\bibitem{acemoglu2012network}
Daron Acemoglu, Vasco~M Carvalho, Asuman Ozdaglar, and Alireza Tahbaz-Salehi.
\newblock Network origins of aggregate fluctuations.
\newblock {\em Econometrica}, 80(5):1977--2016, 2012.

\bibitem{acemoglu2015systemic}
Daron Acemoglu, Asuman Ozdaglar, and Alireza Tahbaz-Salehi.
\newblock Systemic risk and stability in financial networks.
\newblock {\em American Economic Review}, 105(2):564--608, 2015.

\bibitem{amini2016risk}
Hamed Amini and Andreea Minca.
\newblock Risk management for systemic risk: Reserve requirements and fire sale effects.
\newblock {\em Operations Research Letters}, 44(4):538--543, 2016.

\bibitem{anand2015filling}
Kartik Anand, Ben Craig, and Goetz Von~Peter.
\newblock Filling in the blanks: Network structure and interbank contagion.
\newblock {\em Quantitative Finance}, 15(4):625--636, 2015.

\bibitem{arnold2012systemic}
Bruce Arnold, Claudio Borio, Luci Ellis, and Fariborz Moshirian.
\newblock Systemic risk, macroprudential policy frameworks, monitoring financial systems and the evolution of capital adequacy.
\newblock {\em Journal of Banking \& Finance}, 36(12):3125--3132, 2012.

\bibitem{banerjee2018uniqueness}
Tathagata Banerjee, Alex Bernstein, and Zachary Feinstein.
\newblock Dynamic clearing and contagion in financial networks.
\newblock {\em arXiv:1801.02091}, 2018.

\bibitem{banerjee2019impact}
Tathagata Banerjee and Zachary Feinstein.
\newblock Impact of contingent payments on systemic risk in financial networks.
\newblock {\em Mathematics and Financial Economics}, 13(4):617--636, 2019.

\bibitem{baqaee2019macroeconomic}
David~Rezza Baqaee and Emmanuel Farhi.
\newblock Macroeconomic effects of aggregate supply relations.
\newblock {\em Annual Review of Economics}, 11:471--497, 2019.

\bibitem{bardoscia2019full}
Marco Bardoscia, Gerardo Ferrara, Nicholas Vause, and Michael Yoganayagam.
\newblock Full payment algorithm.
\newblock {\em Available at SSRN}, 2019.

\bibitem{barratt2020multi}
Shane Barratt and Stephen Boyd.
\newblock Multi-period liability clearing via convex optimal control.
\newblock {\em arXiv:2005.09066}, 2020.

\bibitem{barucca2016network}
Paolo Barucca, Marco Bardoscia, Fabio Caccioli, Marco D'Errico, and Gabriele Visentin.
\newblock Network valuation in financial systems.
\newblock {\em Mathematical Finance}, 2016.

\bibitem{biais2012dynamic}
Bruno Biais, Thomas Mariotti, Jean-Charles Rochet, and St{\'e}phane Villeneuve.
\newblock Dynamic security design: Convergence to continuous time and asset pricing implications.
\newblock {\em The Review of Economic Studies}, 74(2):345--390, 2012.

\bibitem{bourget2023geometry}
Antoine Bourget.
\newblock The geometry of quivers.
\newblock {\em Phys. Sci. Forum}, 5(1):42, 2023.

\bibitem{caccioli2014stability}
Fabio Caccioli, Munik Shrestha, Cristopher Moore, and J~Doyne Farmer.
\newblock Stability analysis of financial contagion due to overlapping portfolios.
\newblock {\em Journal of Banking \& Finance}, 46:233--245, 2014.

\bibitem{capponi2015systemic}
Agostino Capponi and Peng-Chu Chen.
\newblock Systemic risk mitigation in financial networks.
\newblock {\em Journal of Economic Dynamics and Control}, 58:152--166, 2015.

\bibitem{cifuentes2005liquidity}
Rodrigo Cifuentes, Gianluigi Ferrucci, and Hyun~Song Shin.
\newblock Liquidity risk and contagion.
\newblock {\em Journal of the European Economic Association}, 3(2-3):556--566, 2005.

\bibitem{cont2013network}
Rama Cont, Amal Moussa, and Edson~Bastos Santos.
\newblock Network structure and systemic risk in banking systems.
\newblock pages 327--368, 2013.

\bibitem{cont2019monitoring}
Rama Cont and Eric Schaanning.
\newblock Monitoring indirect contagion.
\newblock {\em Journal of Banking \& Finance}, 104:85--102, 2019.

\bibitem{deMarco2020fictitious}
Giuseppe De~Marco, Chiara Donnini, Federica Gioia, and Francesca Perla.
\newblock On the fictitious default algorithm in fuzzy financial networks.
\newblock {\em International Journal of Approximate Reasoning}, 121:1--17, 2020.

\bibitem{demange2018multilayer}
Gabrielle Demange.
\newblock Multilayer networks in economics.
\newblock {\em Annual Review of Economics}, 10:435--461, 2018.

\bibitem{dijkstra1980termination}
Edsger~W Dijkstra and CS~Scholten.
\newblock Termination detection for diffusing computations.
\newblock {\em Information Processing Letters}, 11(1):1--4, 1980.

\bibitem{duffie2015systemic}
Darrell Duffie.
\newblock Systemic risk exposures: a 10-by-10-by-10 approach.
\newblock pages 47--56, 2015.

\bibitem{eisenberg2001systemic}
Larry Eisenberg and Thomas~H Noe.
\newblock Systemic risk in financial systems.
\newblock {\em Management Science}, 47(2):236--249, 2001.

\bibitem{elliott2014financial}
Matthew Elliott, Benjamin Golub, and Matthew~O Jackson.
\newblock Financial networks and contagion.
\newblock {\em American Economic Review}, 104(10):3115--3153, 2014.

\bibitem{elsinger2009financial}
Helmut Elsinger.
\newblock Financial networks, cross holdings, and limited liability.
\newblock {\em Working Papers, Oesterreichische Nationalbank (Austrian Central Bank)}, (156), 2009.

\bibitem{elsinger2006risk}
Helmut Elsinger, Alfred Lehar, and Martin Summer.
\newblock Risk assessment for banking systems.
\newblock {\em Management Science}, 52(9):1301--1314, 2006.

\bibitem{feinstein2019obligations}
Zachary Feinstein.
\newblock Obligations with physical delivery in a multilayered financial network.
\newblock {\em SIAM Journal on Financial Mathematics}, 10(4):877--906, 2019.

\bibitem{fischer2014no}
Marcel Fischer and Christian Pickard.
\newblock No bank is an island: The interplay of capital, liquidity and seniority in financial networks.
\newblock {\em Available at SSRN}, 2014.

\bibitem{furfine2003interbank}
Craig~H Furfine.
\newblock Interbank exposures: Quantifying the risk of contagion.
\newblock {\em Journal of Money, Credit and Banking}, pages 111--128, 2003.

\bibitem{gai2011complexity}
Prasanna Gai, Andrew Haldane, and Sujit Kapadia.
\newblock Complexity, concentration and contagion.
\newblock {\em Journal of Monetary Economics}, 58(5):453--470, 2011.

\bibitem{gandy2017bayesian}
Axel Gandy and Luitgard~AM Veraart.
\newblock A bayesian methodology for systemic risk assessment in financial networks.
\newblock {\em Management Science}, 63(12):4428--4446, 2017.

\bibitem{Ghrist_2022}
Robert Ghrist and Hans Riess.
\newblock Cellular sheaves of lattices and the tarski {L}aplacian.
\newblock {\em Homology, Homotopy and Applications}, 24(1):325--345, 2022.

\bibitem{glasserman2015likely}
Paul Glasserman and H~Peyton Young.
\newblock How likely is contagion in financial networks?
\newblock {\em Journal of Banking \& Finance}, 50:383--399, 2015.

\bibitem{glasserman2016contagion}
Paul Glasserman and H~Peyton Young.
\newblock Contagion in financial networks.
\newblock {\em Journal of Economic Literature}, 54(3):779--831, 2016.

\bibitem{hansen2019}
Jakob Hansen and Robert Ghrist.
\newblock Toward a spectral theory of cellular sheaves.
\newblock {\em Journal of Applied and Computational Topology}, 3:315--358, 2019.

\bibitem{hansen2020}
Jakob Hansen and Robert Ghrist.
\newblock Opinion dynamics on discourse sheaves.
\newblock {\em SIAM Journal on Applied Mathematics}, 81(5):2033--2060, 2021.

\bibitem{jackson2015networks}
Matthew~O Jackson, Brian~W Rogers, and Yves Zenou.
\newblock Networks and the macroeconomy: An empirical exploration.
\newblock {\em Annual Review of Economics}, 7(1):468--486, 2015.

\bibitem{keller2017quiver}
Bernhard Keller.
\newblock Quiver mutation and combinatorial {DT}-invariants.
\newblock {\em arXiv:1709.03143}, 2017.

\bibitem{kusnetsov2019interbank}
Michael Kusnetsov and Luitgard~AM Veraart.
\newblock Interbank clearing in financial networks with multiple maturities.
\newblock {\em SIAM Journal on Financial Mathematics}, 10(1):37--67, 2019.

\bibitem{mistrulli2011assessing}
Paolo~Emilio Mistrulli.
\newblock Assessing financial contagion in the interbank market: Maximum entropy versus observed interbank lending patterns.
\newblock {\em Journal of Banking \& Finance}, 35(5):1114--1127, 2011.

\bibitem{rogers2013failure}
L.~C.~G. Rogers and L.~A.~M. Veraart.
\newblock Failure and rescue in an interbank network.
\newblock {\em Management Science}, 59(4):882--898, 2013.

\bibitem{sumray2024quiver}
Otto Sumray, Heather~A. Harrington, and Vidit Nanda.
\newblock Quiver laplacians and feature selection.
\newblock {\em arXiv:2404.06993}, 2024.

\bibitem{upper2011simulation}
Christian Upper.
\newblock Simulation methods to assess the danger of contagion in interbank markets.
\newblock {\em Journal of Financial Stability}, 7(3):111--125, 2011.

\bibitem{Yu2024}
Lu~Yu.
\newblock Order-theoretical fixed point theorems for correspondences and application in game theory.
\newblock {\em arXiv preprint}, July 2024.

\bibitem{Zhou1994}
Lin Zhou.
\newblock The set of nash equilibria of a supermodular game is a complete lattice.
\newblock {\em Games and Economic Behavior}, 7(2):295--300, 1994.

\end{thebibliography}


\appendix

\section{Lattice Theory Foundations}
\label{sec:foundations}

The order-theoretic foundations of our work rest on lattices -- algebraic structures that formalize the notion of ordering and optimization. We begin with the basic definitions.

A \emph{partially ordered set} (or \emph{poset}) $(P,\leq)$ consists of a set $P$ with a binary relation $\leq$ that is reflexive, antisymmetric, and transitive. For elements $a,b$ in a poset, their \emph{least upper bound} (if it exists) is an element $c$ satisfying $a \leq c$, $b \leq c$, and if $a \leq d$ and $b \leq d$ then $c \leq d$. Their \emph{greatest lower bound} is defined dually.

A \emph{lattice} $(L,\leq)$ is a poset in which every pair of elements has both a least upper bound (denoted $a \vee b$ and called the \emph{join}) and a greatest lower bound (denoted $a \wedge b$ and called the \emph{meet}). The operations $\vee$ and $\wedge$ satisfy:
\[
\begin{aligned}
a \vee b &= b \vee a & \text{(commutativity)} \\
a \vee (b \vee c) &= (a \vee b) \vee c & \text{(associativity)} \\
a \vee (a \wedge b) &= a & \text{(absorption)} \\
a \vee a &= a & \text{(idempotence)}
\end{aligned}
\]
with dual properties holding for $\wedge$.

A lattice $L$ is \emph{complete} if every subset $S \subseteq L$ has both a least upper bound $\bigvee S$ and a greatest lower bound $\bigwedge S$ in $L$. Any closed interval $[a,b]$ in $\mathbb{R}$ is complete, as is the power set of any set under inclusion. The product of complete lattices is complete under the component-wise order -- a fact essential to our construction of network state spaces.

Given lattices $L$ and $M$, a function $f: L \to M$ is \emph{monotone} if $x \leq y$ implies $f(x) \leq f(y)$ for all $x,y \in L$. Monotone maps are the most general structure-preserving maps between lattices; they preserve the order but not necessarily the lattice operations. A stronger notion is that of a \emph{join-preserving} map, where $f(x \vee y) = f(x) \vee f(y)$ for all $x,y \in L$. While join-preservation appears naturally in some contexts (for instance, in valuations of financial portfolios), monotonicity alone suffices for many applications and provides greater modeling flexibility.

The lattices arising in our network models are typically products of real intervals (modeling payments or resources) or more complex constructions modeling obligations and constraints. The crucial feature is that the order structure captures meaningful comparisons between states, while monotone maps model how local changes propagate through the network while respecting these comparisons.

\section{Tarski Fixed Point Theorem}
\label{sec:tarski}

The cornerstone of our analysis is Tarski's fixed point theorem, which guarantees the existence of fixed points for monotone maps on complete lattices and provides structure to the set of such fixed points.

Let $(L,\leq)$ be a complete lattice and $f: L \to L$ a monotone function. The set of \emph{fixed points} of $f$ is $\Fix(f) = \{x \in L : f(x) = x\}$. An element $x \in L$ is called an \emph{upper bound} of $f$ if $x \geq f(x)$, and a \emph{lower bound} if $x \leq f(x)$.

\begin{theorem}[Tarski]
Let $(L,\leq)$ be a complete lattice and $f: L \to L$ be monotone. Then:
\begin{enumerate}
\item $\Fix(f)$ is nonempty
\item $\Fix(f)$ forms a complete lattice under the induced order from $L$
\item The least fixed point of $f$ is $\bigwedge\{x \in L : f(x) \leq x\}$
\item The greatest fixed point of $f$ is $\bigvee\{x \in L : x \leq f(x)\}$
\end{enumerate}
\end{theorem}

{\em Proof:} Let $U = \{x \in L : x \geq f(x)\}$ be the set of upper bounds of $f$. Since $L$ is complete, $U$ has a greatest lower bound $a = \bigwedge U$. We claim $a$ is a fixed point.

First, we establish $f(a) \leq a$. Since $a$ is a lower bound of $U$, for any $x \in U$ we have $a \leq x$. By monotonicity of $f$, this implies $f(a) \leq f(x)$. For any $x \in U$, we have $f(x) \leq x$ by definition of $U$. Combining these inequalities gives $f(a) \leq f(x) \leq x$ for all $x \in U$. Thus, $f(a)$ is also a lower bound of $U$. Since $a$ is the greatest lower bound of $U$, we must have $f(a) \leq a$.

Next, we show $a \leq f(a)$. From the first part, we know $f(a) \leq a$. Applying $f$ to both sides and using monotonicity, we get $f(f(a)) \leq f(a)$. This means $f(a) \in U$ by definition of $U$. Since $a$ is a lower bound of $U$ and $f(a) \in U$, we must have $a \leq f(a)$.
Combining these inequalities, we obtain $f(a) = a$, proving that $a$ is a fixed point of $f$.

For completeness of $\Fix(f)$, let $S \subseteq \Fix(f)$ be any subset. Let $b = \bigvee S$ in $L$. Then for any $s \in S$, $s \leq b$ implies $f(s) = s \leq f(b)$ by monotonicity. Thus $b = \bigvee S \leq f(b)$. A dual argument shows $f(b) \leq b$, so $b \in \Fix(f)$. Similarly, $\bigwedge S \in \Fix(f)$. 
\qed

The constructive aspects of Tarski's theorem yield important algorithmic insights. Starting from any lower bound $x_0 \leq f(x_0)$, the sequence $x_n = f^n(x_0)$ is monotone increasing and converges to a fixed point. Dually, iteration from an upper bound yields a decreasing sequence converging to a fixed point. When $L$ has finite height, these sequences terminate in finitely many steps.

The lattice structure of $\Fix(f)$ has vital implications for network equilibria. When modeling payment flows or resource allocation, the least fixed point often represents a worst-case scenario where defaults cascade maximally, while the greatest fixed point represents an optimal clearing where obligations are met to the greatest extent possible. The existence of these extremal fixed points provides bounds on all possible system behaviors.


\section{Chain-Complete Lattices}
\label{sec:chains}

Complete lattices play a central role in our framework, but many applications naturally lead to partially ordered sets that satisfy weaker completeness conditions. This appendix provides the necessary background on chain-complete lattices and the corresponding fixed point theorems that extend our results to these more general settings.

\begin{definition}[Chain]
A \emph{chain} in a partially ordered set $(P, \leq)$ is a subset $C \subseteq P$ such that for any $x, y \in C$, either $x \leq y$ or $y \leq x$. In other words, a chain is a totally ordered subset of $P$.
\end{definition}

\begin{definition}[Height]
The \emph{height} of a chain $C$ in a partially ordered set is the number of elements in the chain minus one (representing the number of strict inequalities in the chain). The \emph{height} of a partially ordered set $(P, \leq)$ is the supremum of the heights of all chains in $P$. For a finite lattice, the height represents the length of the longest chain from the bottom element to the top element.
\end{definition}

\begin{definition}[Chain-Completeness]
A partially ordered set $(P, \leq)$ is called \emph{chain-complete} if every nonempty chain $C \subseteq P$ has a supremum $\sup C$ in $P$. A poset is \emph{dual chain-complete} if every nonempty chain has an infimum $\inf C$ in $P$. A poset is \emph{bi-chain-complete} if it is both chain-complete and dual chain-complete.
\end{definition}

Every complete lattice is trivially bi-chain-complete, but the converse does not hold. Many naturally occurring partially ordered sets in applications are chain-complete without being complete lattices.

\begin{example}
The set $\mathcal{F}$ of all continuous functions $f: [0,1] \to \mathbb{R}$ with pointwise ordering $(f \leq g$ if $f(x) \leq g(x)$ for all $x \in [0,1])$ is chain-complete but not a complete lattice. Given any chain of continuous functions, their supremum is again continuous, but the supremum of an uncountable set of continuous functions may not be continuous.
\end{example}

\begin{definition}[Scott Continuity]
Let $(P, \leq)$ and $(Q, \leq)$ be posets. A function $f: P \to Q$ is \emph{Scott-continuous} if for every nonempty chain $C \subseteq P$ with a supremum in $P$, the supremum of $f(C) = \{f(x) : x \in C\}$ exists in $Q$ and:
\[
f(\sup C) = \sup f(C)
\]
More generally, Scott-continuity refers to the preservation of suprema of directed sets, where a directed set is a non-empty set in which every finite subset has an upper bound in the set.
\end{definition}

The Kleene-Tarski fixed point theorem extends Tarski's result to chain-complete posets, provided we strengthen monotonicity to Scott-continuity:

\begin{theorem}[Kleene-Tarski Fixed Point Theorem]
\label{thm:chainTarski}
Let $(P, \leq)$ be a chain-complete poset with a least element $\bot$, and let $f: P \to P$ be a Scott-continuous function. Then:
\begin{enumerate}
\item $f$ has a least fixed point $\mu f$
\item This least fixed point can be constructed as $\mu f = \sup \{f^n(\bot) : n \geq 0\}$, where $f^n$ denotes the $n$-fold composition of $f$
\item If $f$ is also monotone and $P$ is bi-chain-complete with a greatest element $\top$, then $f$ has a greatest fixed point $\nu f = \inf \{f^n(\top) : n \geq 0\}$
\end{enumerate}
\end{theorem}

{\em Proof:} 
Define the sequence $\{x_n\}_{n \geq 0}$ by $x_0 = \bot$ and $x_{n+1} = f(x_n)$ for $n \geq 0$. By induction, since $\bot \leq f(\bot)$ and $f$ is monotone, we have $x_n \leq x_{n+1}$ for all $n \geq 0$. Thus $\{x_n\}$ forms a chain in $P$.

Since $P$ is chain-complete, the supremum $x = \sup \{x_n : n \geq 0\}$ exists in $P$. By Scott-continuity of $f$:
\[
f(x) = f(\sup \{x_n : n \geq 0\}) = \sup \{f(x_n) : n \geq 0\} = \sup \{x_{n+1} : n \geq 0\} = x
\]
Thus $x$ is a fixed point of $f$.

To see that $x$ is the least fixed point, let $y$ be any fixed point of $f$. By induction, $x_0 = \bot \leq y$, and if $x_n \leq y$, then $x_{n+1} = f(x_n) \leq f(y) = y$. Thus $x_n \leq y$ for all $n \geq 0$, which implies $x = \sup \{x_n\} \leq y$.

The proof for the greatest fixed point is dual when $P$ has a greatest element $\top$.
\qed

The Kleene-Tarski theorem not only guarantees the existence of fixed points but also provides a constructive method for finding them through iterative approximation, starting from the least or greatest element. This constructive aspect is particularly valuable for computational applications where fixed points must be computed explicitly.

\begin{definition}[Directed Set]
A nonempty subset $D$ of a poset $(P, \leq)$ is \emph{directed} if for any $x, y \in D$, there exists $z \in D$ such that $x \leq z$ and $y \leq z$. Dually, $D$ is \emph{filtered} if for any $x, y \in D$, there exists $z \in D$ such that $z \leq x$ and $z \leq y$.
\end{definition}

\begin{definition}[Directed-Complete Partial Order]
A partially ordered set $(P, \leq)$ is a \emph{directed-complete partial order} (dcpo) if every directed subset $D \subseteq P$ has a supremum $\sup D$ in $P$.
\end{definition}

Chain-completeness and directed-completeness are related:

\begin{proposition}
Every dcpo is chain-complete. If a poset is chain-complete and has binary joins (i.e., any two elements have a least upper bound), then it is a dcpo.
\end{proposition}

For monotone functions between posets, we have the following fixed point theorem due to Pataraia:

\begin{theorem}[Pataraia's Fixed Point Theorem]
Let $(P, \leq)$ be a dcpo with a least element $\bot$, and let $f: P \to P$ be a monotone function. Then $f$ has a least fixed point.
\end{theorem}

These generalizations allow our framework to accommodate a wider range of application domains where the complete lattice structure may be too restrictive, while still providing the fixed-point guarantees essential for our analysis of clearing sections in network models.

\section{Residuated Lattices}
\label{sec:residuation}

Residuated lattices provide a rich algebraic structure for modeling logical operations in various non-classical logics, particularly fuzzy logics. This appendix presents the basic theory of residuated lattices and shows how they naturally arise in applications involving qualitative or fuzzy attributes.

\begin{definition}[Residuated Lattice]
A \emph{residuated lattice} is a structure $(\mathcal{R}, \wedge, \vee, \otimes, \rightarrow, 0, 1)$ where:
\begin{enumerate}
\item $(\mathcal{R}, \wedge, \vee, 0, 1)$ is a bounded lattice with least element 0 and greatest element 1
\item $(\mathcal{R}, \otimes, 1)$ is a commutative monoid with identity element 1
\item The binary operations $\otimes$ (called \emph{multiplication}) and $\rightarrow$ (called \emph{residuum}) form an adjoint pair, meaning that for all $x, y, z \in \mathcal{R}$:
\[
x \otimes y \leq z \text{ if and only if } x \leq (y \rightarrow z)
\]
\end{enumerate}
\end{definition}

The adjoint relationship between $\otimes$ and $\rightarrow$ generalizes the classical relationship between conjunction and implication in Boolean logic. In residuated lattices, however, truth values can be drawn from arbitrary bounded lattices rather than just the binary set $\{0,1\}$.

\begin{proposition}
In a residuated lattice, the following properties hold for all $x, y, z \in \mathcal{R}$:
\begin{enumerate}
\item $\otimes$ is monotone in both arguments: if $x \leq y$, then $x \otimes z \leq y \otimes z$
\item $\rightarrow$ is monotone in the second argument: if $y \leq z$, then $x \rightarrow y \leq x \rightarrow z$
\item $\rightarrow$ is antitone in the first argument: if $x \leq y$, then $y \rightarrow z \leq x \rightarrow z$
\item $x \otimes (x \rightarrow y) \leq y$
\item $x \leq (y \rightarrow (x \otimes y))$
\item $x \rightarrow x = 1$
\item $x \otimes 1 = x$
\item $1 \rightarrow x = x$
\end{enumerate}
\end{proposition}

{\em Proof:} These properties follow directly from the definition of residuated lattices and the adjoint relationship between $\otimes$ and $\rightarrow$. For instance, to show (1), assume $x \leq y$. Since $y \otimes z \leq y \otimes z$, by the adjoint property we have $y \leq (z \rightarrow (y \otimes z))$. Since $x \leq y$, we get $x \leq (z \rightarrow (y \otimes z))$, which by the adjoint property implies $x \otimes z \leq y \otimes z$.

Properties (4) through (8) can be proved similarly using the adjoint relationship and the properties of bounded lattices and commutative monoids.
\qed

Residuated lattices provide an algebraic framework for various non-classical logics, particularly fuzzy logics. Different choices of operations $\otimes$ and $\rightarrow$ correspond to different logical systems:

Different choices of the operations $\otimes$ and $\rightarrow$ on the interval $[0,1]$ yield important fuzzy logics:

\begin{itemize}
\item \textbf{G\"odel logic}: $x \otimes y = \min(x, y)$ and $x \rightarrow y = 1$ if $x \leq y$, otherwise $x \rightarrow y = y$. This logic takes a conservative approach where conjunction is the minimum value, appropriate when the weakest component determines overall strength.

\item \textbf{\L{}ukasiewicz logic}: $x \otimes y = \max(0, x + y - 1)$ and $x \rightarrow y = \min(1, 1 - x + y)$. This logic allows for accumulation and cancellation effects, with partial truths combining additively.

\item \textbf{Product logic}: $x \otimes y = x \cdot y$ (ordinary product) and $x \rightarrow y = 1$ if $x \leq y$, otherwise $x \rightarrow y = y/x$. This logic models independent contributions multiplicatively, suitable for combining probabilities or evidence.
\end{itemize}

Each system models different aspects of fuzzy reasoning and applies to different application domains depending on how truth values should combine.

Each of these logics satisfies different properties. For instance, in G\"odel logic, we have idempotence of conjunction ($x \otimes x = x$), while in \L{}ukasiewicz logic, we can have $x \otimes x < x$ for $x > 0$. These differences lead to different behaviors when modeling aspects like evidence accumulation, logical consistency, or degree of satisfaction.

Residuated lattices can be extended to handle more complex logical operations:

\begin{definition}
In a residuated lattice $(\mathcal{R}, \wedge, \vee, \otimes, \rightarrow, 0, 1)$, we define:
\begin{enumerate}
\item Negation: $\neg x = x \rightarrow 0$
\item Biresiduum or equivalence: $x \leftrightarrow y = (x \rightarrow y) \wedge (y \rightarrow x)$
\end{enumerate}
\end{definition}

Unlike classical logic, negation in residuated lattices might not satisfy the law of excluded middle ($x \vee \neg x = 1$) or the law of non-contradiction ($x \wedge \neg x = 0$), which makes them suitable for modeling various forms of uncertainty, partiality, or vagueness.

The algebraic structure of residuated lattices provides a rigorous foundation for reasoning about graded or partial truth values, making them ideal for applications where attributes are not simply present or absent but can be satisfied to varying degrees.

\end{document}